\documentclass[12pt,preprint]{aastex}

\lefthead{Chou et al.}
\righthead{Abundances of Sagittarius
Stream Stars}

\begin{document}
\title{A Two Micron All-Sky Survey View of the Sagittarius Dwarf Galaxy:\\
VI. s-Process and Titanium Abundance Variations Along the
Sagittarius Stream}

\author{Mei-Yin Chou\altaffilmark{1},
 Katia Cunha\altaffilmark{2,3},
Steven R. Majewski\altaffilmark{1},
Verne V. Smith\altaffilmark{2}, \\
Richard J. Patterson\altaffilmark{1}, David
Mart{\'i}nez-Delgado\altaffilmark{4} and Doug
Geisler\altaffilmark{5} }

\altaffiltext{1}{Dept. of Astronomy, University of Virginia,
Charlottesville, VA 22904-4325 (mc6ss, srm4n, rjp0i@virginia.edu)}

\altaffiltext{2}{National Optical Astronomy Observatories, PO Box
26732, Tucson, AZ 85726 (cunha, vsmith@noao.edu)}

\altaffiltext{3}{On leave from Observatorio Nacional, Rio de
Janeiro, Brazil}

\altaffiltext{4}{Instituto de Astrofisica de Canarias, La Laguna,
Spain (ddelgado@iac.es)}

\altaffiltext{5}{Departamento de Astronomia, Universidad de
Concepci\'{o}n, Casilla 160-C, Concepci\'{o}n, Chile
(dgeisler@astro-udec.cl)}

\begin{abstract}

We present high-resolution spectroscopic measurements of the
abundances of the $\alpha$ element titanium (Ti) and s-process
elements yttrium (Y) and lanthanum (La) for 59 candidate M giant
members of the Sagittarius (Sgr) dwarf spheroidal (dSph) + tidal
tail system pre-selected on the basis of position and radial
velocity.  As expected, the majority of these stars show peculiar
abundance patterns compared to those of nominal Milky Way stars,
but as a group the stars form a coherent picture of chemical
enrichment of the Sgr dSph from [Fe/H] = -1.4 to solar abundance.
This sample of spectra provides the largest number of Ti, La and Y
abundances yet measured for a dSph, and spans metallicities not
typically probed by studies of the other, generally more
metal-poor Milky Way (MW) satellites. On the other hand, the
overall [Ti/Fe], [Y/Fe], [La/Fe] and [La/Y] patterns with [Fe/H]
of the Sgr stream plus Sgr core do, for the most part, resemble
those seen in the Large Magellanic Cloud (LMC) and other dSphs,
only shifted by $\Delta$[Fe/H]$\sim$+0.4 from the LMC and by
$\sim$+1 dex from the other dSphs; these relative shifts reflect
the faster and/or more efficient chemical evolution of Sgr
compared to the other satellites, and show that Sgr has had an
enrichment history more like the LMC than the other dSphs. By
tracking the evolution of the abundance patterns along the Sgr
stream we can follow the time variation of the chemical make-up of
dSph stars donated to the Galactic halo by Sgr.  This evolution
demonstrates that while the bulk of the stars currently in the Sgr
dSph are quite unlike those of the Galactic halo, an increasing
number of stars farther along the Sgr stream have abundances like
Milky Way halo stars, a trend that shows clearly how the Galactic
halo could have been contributed by present day satellite galaxies
even if the {\it present} chemistry of those satellites is now
different from typical halo field stars. Finally, we analyze the
chemical abundances of a moving group of M giants among the Sgr
leading arm stars at the North Galactic Cap, but having radial
velocities unlike the infalling Sgr leading arm debris there.
Through use of ``chemical fingerprinting'', we conclude that these
mostly receding northern hemisphere M giants also are Sgr stars,
likely {\it trailing arm} debris overlapping the Sgr leading arm
in the north.

\end{abstract}


\section{Introduction}

Though it is now clear that accretion of dwarf galaxies likely
played a prominent role in creating the Milky Way's (MW) stellar
halo (Searle \& Zinn 1978), with strong observational evidence
(e.g., Majewski 1993; Majewski, Munn \& Hawley 1996; Bell et al.
2008) and a theoretical backing by $\Lambda$CDM models (e.g.,
Bullock \& Johnston 2005; Robertson et al. 2005; Abadi et al.\
2006; Font et al.\ 2006a), it is often highlighted that the
chemical abundance patterns of current MW satellites are rather
different than those of halo field stars, which typically show
significantly higher [$\alpha$/Fe] than MW dSph stars at the same
[Fe/H] (e.g., Fulbright 2002; Shetrone et al.\ 2003; Tolstoy et
al.\ 2003; Venn et al.\ 2004; Geisler et al.\ 2005). The reason
for these differences remains a matter of speculation. One
interpretation is that the dwarf systems that contributed the bulk
of the halo were more massive, Magellanic Cloud-sized systems that
were accreted and destroyed very early on, so that the chemistry
of the accreted stars was necessarily dominated only by enrichment
from Type II supernovae (SN II)(Robertson et al.\ 2005; Font et
al.\ 2006a). Alternatively (or in addition), if current dwarf
satellites have experienced both prolonged chemical evolution and
tidal disruption, this will naturally lead to evolution in the
types of stars that satellites contribute to the Galactic halo
(Majewski et al.\ 2000, 2002).

In principle, one can test the bridge from dwarf galaxy chemistry
to halo star chemistry {\it directly} if one can identify the
stars already contributed by the dwarf galaxy to the halo. Perhaps
the easiest way to do this is by exploring elemental abundances
along the tidal tails of disrupting dwarf galaxies. Presently, the
best known example of a tidally disrupting Milky Way satellite is
the Sagittarius (Sgr) dwarf spheroidal (dSph) galaxy, a system
that is especially interesting for understanding the issues raised
above because the proximity of its core and tails make it
particularly accessible to high resolution spectroscopic study,
and because the tails produced by its steady assimilation into the
MW have been tracked over a substantial length (e.g., Ibata et
al.\ 2001; Majewski et al.\ 2003, hereafter Paper I; Belokurov et
al.\ 2007) corresponding to several gigayears of tidal disruption
(Law et al.\ 2005, ``Paper IV'' hereafter). Thus the extensive Sgr
tidal stream can provide a key link between dwarf galaxies, tidal
disruption and capture, and the chemical evolution and origin of
the MW halo.

Exploration of the chemistry of any dwarf galaxy can be used to
study star formation environments that are quite distinct from
those in the MW, but Sgr provides an interesting case study for
chemical evolution in its own right. It is the most luminous and
massive of the current MW dSph systems, and has a history
punctuated by a number of star formation episodes (Sarajedini \&
Layden 1995; Layden \& Sarajedini 2000; Siegel et al.\ 2007). This
active star formation history quickly elevated Sgr's mean
metallicity to almost solar by about 5 Gyr ago (Bellazzini et al.\
2006; Siegel et al.\ 2007), making it the dSph with the most
metal-rich stellar populations known associated with the Milky
Way. Because Sgr has been tidally disrupting for at least 2.5-3.0
Gyr (Paper IV), in principle this means that Sgr could have
contributed stars with a wide metallicity range to the Galactic
halo.

This has been confirmed by the recent analysis of Chou et al.
(2007, hereafter Paper V), who measured [Fe/H] ranging from -1.4
dex to 0.0 dex for stars identified spatially, dynamically and by
spectral type (i.e. the M giants characteristic of the Sgr dSph)
with the Sgr stream. Interestingly, the Paper V analysis revealed
a strong metallicity gradient along the Sgr stream --- proof of a
time dependence in the enrichment level of the stars donated to
the halo by Sgr, and evidence for preferential tidal stripping of
metal poor stars, which has led to divergent metallicity
distribution functions (MDFs) between lost and retained Sgr stars.
A similar metallicity gradient in the Sgr leading and trailing
streams has also been seen by the high resolution analysis of M
giants by Monaco et al. (2007). Because stars are typically
stripped from the outer parts of dSphs, the changing MDF along the
Sgr stream suggests a strong metallicity gradient within the
original Sgr system.

Sgr is also chemically interesting because its abundance patterns
reveal that while it is undersolar and depleted with respect to
the MW pattern for its $\alpha$-, odd-$Z$ and iron peak-elements
among its [Fe/H] $\gtrsim$ -1 populations, Sgr's more metal-poor
populations seem to have [$\alpha$/Fe] that {\it do} resemble
those of the Galactic halo (Smecker-Hane \& McWilliam 2002,
Bonifacio et al.\ 2004, Monaco et al.\ 2005b, Sbordone et al.\
2007). Presently these findings are based on only about 6 Sgr
stars having both [Fe/H] $< -1$ and measured [$\alpha$/Fe] from
all of the above studies, as well as a handful of stars from M54
(possibly the nucleus of Sgr --- see discussions by Ibata et al.\
1994; Sarajedini et al.\ 1995; Da Costa \& Armandroff 1995;
Bassino \& Muzzio 1995; Layden \& Sarajedini 2000; Paper I, but
also cf. Monaco et al.\ 2005a) in the study by Brown et al.\
(1999). However, if true, the trend may not be unique to Sgr:
Abundance patterns (e.g., [$\alpha$/Fe]) for some of the very most
metal-poor stars in other dSphs also seem to overlap those of halo
stars of the same metallicity (Shetrone et al.\ 2003; Geisler et
al.\ 2005; Tolstoy 2005). Thus, if {\it these} dSph systems
experienced tidal disruption, the few very metal poor stars they
now hold with MW-like abundance patterns may only represent the
residue of a formerly much larger metal-poor population that may
have been predominantly stripped from the satellites over their
lifetime (see discussion of this in the case of the Carina dSph in
Majewski et al. 2002 and Mu\~noz et al. 2006). Such a scenario
could explain the possible origins of the abundance dichotomy
between the present dSphs and the Galactic halo.

In this paper we continue our detailed exploration of Sgr stream
stars from Paper V with a focus on the chemical trends of the
$\alpha$-element titanium (Ti) and the s-process elements yttrium
(Y) and lanthanum (La) along the stream. Our study is based on the
largest sample of high resolution spectra yet obtained of Sgr
stars, and includes a significant number for stars with [Fe/H] $<
-0.9$ (\S2-3). Collectively, and when combined with additional
data from the literature, these stars present a much more complete
picture of the chemical abundance patterns for Sgr stars,
including the first view of Sgr s-process abundances for [Fe/H]
$<-0.9$. In \S4 we show that, as seen before with smaller samples,
the Sgr [$\alpha$/Fe] abundances are enhanced and MW-like among
the more metal-poor stars, but we also show, for the first time
that this trend also arises among the s-process elements we
explore. Because of the overall variation of the MDF reported in
Paper V, there are increasing numbers of stars with ``MW-like''
abundances with distance along the stream away from the Sgr core.
Thus, the Sgr stream provides a direct connection between the
unusual, non-MW-like abundance patterns in the Sgr core and stars
with halo-like abundance patterns contributed by Sgr several Gyr
ago. In this way, Sgr provides a counter-example of the Robertson
et al. (2005) and Font et al. (2006a) hypothesis that the
$\alpha$-enhanced stars of the halo were deposited there long ago.

In \S 5 we also discuss the
implications of the Sgr abundance patterns for the chemical
evolution of this dSph. By comparison of the Sgr abundance
patterns to those of other MW satellites, we show that Sgr more
closely resembles the Large Magellanic Cloud in its chemical
evolution than the other dSph-type systems.

Finally, throughout this paper we simultaneously analyze the
chemical abundances of a moving group of M giants among the Sgr
leading arm stars at the North Galactic Cap but having different
radial velocities than the infalling Sgr leading arm debris there.
In Paper V we showed these stars to have a very similar MDF to
leading Sgr arm stars stripped several Gyr ago, a result expected
if this moving group were constituted by Sgr stars stripped at
about the same time. In a demonstration of the technique of
``chemical fingerprinting'', we conclude from their peculiar Ti, Y
and La patterns, which also match those of the extreme Sgr leading
arm, that most of these receding northern hemisphere M giants
could represent Sgr {\it trailing arm} stars in the northern
hemisphere.

\section{Observations}

Our analysis here makes use of the same spectra described in Paper
V for six M giant stars in the Sgr core, thirty candidate stars in
the Sgr leading arm north of the Galactic plane, ten in the Sgr
leading arm but south of the Galactic plane, and thirteen stars in
the North Galactic Cap (NGC) moving group, which have velocities
much higher and mostly opposite those of the infalling Sgr leading
arm there. We selected the first three groups of these stars to be
likely members of the Sgr stream not only by their spatial
distribution, but also based on their radial velocities, which are
appropriate for the Sgr stream at these positions based on Sgr
debris models (Paper IV; Figs.\ 1 and 2 in Paper V). The Sgr core,
leading arm north and leading arm south groups represent a
dynamical sequence from possibly still bound stars to those stars
stripped from Sgr several Gyr ago, respectively. We have argued in
Paper V, and will again here (\S 4), that the NGC moving group
stars look to be former Sgr members with chemistry resembling the
leading arm south stars; given their positions, velocities and
inferred dynamical age, these are most probably Sgr trailing arm
stars in the northern hemisphere.

The stars in our sample were observed with the 6.5-m Clay
telescope at Las Campanas Observatory and the MIKE spectrograph at
$R\sim$19,000, the 4-m Mayall telescope echelle at $R\sim$35,000
at Kitt Peak, and the 3.5-m TNG telescope and SARG spectrograph
operating at $R\sim$46,000 in the Canary Islands \footnote{Based
on observations made with the Italian Telescopio Nazionale Galileo
(TNG) operated on the island of La Palma by the Fundacion Galileo
Galilei of the INAF (Istituto Nazionale di Astrofisica) at the
Spanish Observatorio del Roque de los Muchachos of the Instituto
de Astrofisica de Canarias.}. The data reduction followed standard
procedures and is described in Paper V. Figure 1 shows examples of
portions of the spectra from each instrument including the two Ti
lines used for our analysis here. Further details of the
observations and the positions, photometry and velocities for the
program stars can be found in Paper V (with the latter data
summarized in Table 1 of that paper).

\section{Derivation of Abundances}

The required input parameters for the abundance analysis are
effective temperature $T_{\rm eff}$, surface gravity (usually
parameterized as log $g$), and metallicity. The details on the
determination of the effective temperature, surface gravity, and
iron abundances for our target stars are given in Paper V. The
model atmospheres adopted in the analysis were interpolated from
Kurucz (1994) grids\footnote{From
http://kurucz.harvard.edu/grids.html.} and are the same as those
used in the analysis in Paper V. Abundances in this study were
derived from the LTE code MOOG (Sneden 1973) along with the
adopted model atmospheres from Paper V.

We measure the equivalent widths (EWs) of eleven \ion{Fe}{+1}
lines (used to derive the [Fe/H] presented in Paper V), two
\ion{Ti}{+1} lines and one \ion{Y}{+2} line in a particular part
of the spectrum that is relatively free from TiO and other
molecular contamination and that was
previously investigated by Smith \& Lambert (1985;
1986; 1990 --- hereafter ``S\&L'') in their spectroscopic
exploration of M giants. These particular elements were chosen not
only because they have well-defined, measurable spectral lines,
but also because they show distinct abundance ratios (relative to
Fe) in many dwarf galaxies when compared to the MW. This
circumstance has been found in the core of Sgr itself (Bonifacio
et al.\ 2004; Monaco et al.\ 2005b; Sbordone et al.\ 2007), as
well as in other dSphs (Shetrone et al.\ 2003; Geisler et al.\ 2005) and
the Large Magellanic Cloud (LMC; Smith et al.\ 2002; Johnson et
al.\ 2006; Pomp{\'e}ia et al.\ 2008; Mucciarelli et al.\ 2008).

The $gf$-values for \ion{Ti}{+1} and \ion{Y}{+2} were determined
by measuring their equivalent widths in the solar flux atlas of
Kurucz et al. (1984) and varying the $gf$-values for each line in
order to match the solar titanium and yttrium abundances of
$A$(Ti)=4.90 and $A$(Y)=2.21 (Asplund, Grevesse, \& Sauval 2005);
the adopted solar $gf$-values are listed in Table 1. The measured
EWs of the \ion{Ti}{+1} and \ion{Y}{+2} lines for each of our Sgr
spectra are given in Table 2.  We also include the EW's measured
for several standard stars, which we have analyzed for a
comparison and as a control sample.

We analyzed \ion{La}{+2} via spectral synthesis analysis because
this line is affected by hyperfine splitting. An example of the
spectral synthesis for this line is shown in Figure 2. The
\ion{La}{+2} line we are interested in is from angular momentum
$J=3$ to $J=3$, with nuclear spin $I=7/2$. The lower energy for
this transition is 1016.10 cm$^{-1}$, and the higher energy is
14375.17 cm$^{-1}$. There are nineteen hyperfine splitting lines
for this transition, and the splitting constants are $A=3.38$ and
$B=0.84$ for the lower energy level (Lawler et al. 2001). However,
the $A$ and $B$ constants for the higher level are unknown. We
used various $A$ and $B$ to create a synthetic spectrum and fit it
to a very high-resolution Arcturus spectrum. The best-fit $A$ is
-30. $B$ is a secondary parameter so it actually doesn't affect
the spectrum too much; we have adopted $B=-0.5$ here.

The derived abundance results are summarized in Table 3. For each
star, the columns give the derived effective temperature using the
Houdashelt et al. (2000) color-temperature relation applied to the
2MASS $(J-K_s)_o$ color, and the derived values of the surface
gravity ($\log{g}$), microturbulence ($\xi$), abundance $A$(X),
and abundance ratios [Fe/H] or [X/H] for each element X as well as
the standard deviation in the abundance determinations. The
details for how we derive the atmospheric parameters ($T_{\rm
eff}$, $\log{g}$, [Fe/H], $\xi$) can be found in Paper V. The
standard deviation represents the line to line scatter (for Ti and
Y) and different estimates of the continuum level (for La). We
measured two \ion{Ti}{+1} lines in one order, and two EW
measurements of the same \ion{Y}{+2} line in two adjacent orders.
For \ion{La}{+2}, we have three different abundance measurements
from different continuum level adjustments, and give the resulting
average \ion{La}{+2} abundance and standard deviation of those
values.

We also have measured abundances for the stars Arcturus, $\beta$
Peg, $\beta$ And, $\rho$ Per, and HD 146051, which are nearby K
(Arcturus) and M giants that provide a control sample for our
abundance work. In Table 4 we summarize literature values (S\&L;
McWilliam \& Rich 1994; Smith et al. 2000) for the atmospheric
parameters and abundances for relevant chemical elements of these
control sample stars for comparison to our own derived values (no
such data are available for HD 146051). Because the references
listed adopt somewhat different solar abundances scales, we list
the absolute abundances, $A$(X), rather than abundance ratios in
order to facilitate a comparison. We note that S\&L used the K
giant $\alpha$ Tau as a reference star in their analysis. In order
to compute absolute abundances for the control stars we derive
abundances for $\alpha$ Tau using the equivalent width
measurements in S\&L and the model atmospheres and $gf$ values
adopted in this study. The derived abundances for $\alpha$ Tau are
$A$(Fe)=7.52, $A$(Ti)=4.94 and $A$(Y)=2.32. We used these values
and the relative abundance ratios [X/H] in S\&L to get the
absolute abundances of $\beta$ Peg, $\beta$ And and $\rho$ Per
listed in Table 4. As can be seen, the derived abundances for
Arcturus and $\rho$ Per in this work agree with the literature
values, within the stated errors and with the standard deviation
of the differences less than 0.2 dex. Our derived values of
$\beta$ Peg and $\beta$ And are lower than those in S\&L. This is
due largely to differences in the adopted stellar atmosphere
parameters.

To explore the sensitivity in the derived abundances to changes in
stellar parameters we varied $T_{\rm eff}$, $\log{g}$ and
microturbulence for the control stars and tabulate the abundance
changes in Table 5. This table shows the sensitivities of the
\ion{Fe}{+1}, \ion{Ti}{+1}, \ion{Y}{+2} and \ion{La}{+2} lines
corresponding to changes in $T_{\rm eff}$ by +100 K, $\log{g}$ by
+0.2 dex (where $g$ is measured in ${\rm cm\, s^{-2}}$), and $\xi$
by +0.2 ${\rm km\,s^{-1}}$, respectively.

With these dependencies in hand, we derive the abundances for the
M giant comparison stars adopting the previously published stellar
parameters from S\&L. The resulting abundances are provided as
separate entries in Table 4, and indicate general agreement,
within calculated uncertainties, for A(Fe) values when similar
atmospheric values are adopted for the M giant comparison stars.
This exercise also demonstrates how the derived $A$(Y) values much
more closely match the previously derived values for these stars
when we adopt similar stellar parameters. The lower $A$(Y) we find
in this paper for the control stars arises primarily from our
derivation of lower $\log{g}$ for these stars. As previously
mentioned in Paper V, the S\&L studies predated the availability
of Hipparcos parallaxes, and therefore they adopted higher
$\log{g}$ values from less accurate absolute magnitudes.

Finally, our analysis shows that, in general, we tend to find
$A$(Ti) slightly lower by about 0.2-0.3 dex compared to S\&L, even
when similar atmospheric parameters are adopted for the M giants.
Note however that different sets of Ti lines have been used in the
two studies and that different families of model atmospheres can
also account for some of the abundance differences. In addition,
when comparing the derived abundances with the previously
published results from S\&L one has to keep in mind that S\&L
computed only relative abundances using $\alpha$ Tau as a
reference star with the underlying assumption that its abundance
distribution is approximately solar. Indeed, Kov\'{a}cs (1983)
conducted a detailed absolute abundance analysis for $\alpha$ Tau
and validated this assumption at a level of roughly $\pm 0.2$ dex.
Absolute abundances computed from the relative abundances
published in S\&L will carry the uncertainty in the underlying
$\alpha$ Tau abundance distribution.

Further circumstantial evidence that our Ti abundances are more
reliable comes from a comparison of our derived [Ti/Fe] for the
control stars to the [Ti/Fe] trend of other disk stars, as shown
in the top panel of Figure 3 (where the control sample stars are
shown as brown stars); as may be seen, were we to shift our Ti
abundances several 0.1 dex higher to make them more consistent
with the S\&L values, all five of the control stars (including
HD146051 now) would have [Ti/Fe] above the mean for disk stars of
similar [Fe/H], and in some cases anomalously so. In addition, as
we will show in Figures 3 and 10 below, a several tenths of a dex
offset in $A$(Ti) would make the already high [Ti/Fe] abundances
found for metal poor Sgr stars compared to other dSphs and the LMC
even more extreme.

\section{Derived Abundance Patterns\\ and Variation Along the Sgr Stream}

\subsection{Abundance Differences Compared to Milky Way Field Stars}

To verify whether our sample is indeed dominated by members of the
Sgr stream, rather than random MW field stars, we appeal to the
well-known abundance pattern differences between the MW and dSph
satellites in general, and between the MW and the Sgr dSph in
particular. In principle, one might expect that stars recently
stripped from dSph systems to bear similar peculiar chemical
hallmarks as the stars they left behind in the dSph core. The
promise of ``chemical fingerprinting'' stars to their birth
systems has long been discussed (Freeman \& Bland-Hawthorne 2002;
De Silva et al. 2007), but has yet to be used much in practice. A
side benefit of our study is that it lends itself to a direct test
of the viability of chemical fingerprinting in a well-defined,
fairly controlled context (i.e., testing a sample of stars
specifically selected to have been born in one particular system
-- Sgr --- against contamination from stars from another system
--- the MW).
In addition, in \S 4.2 we {\it apply} chemical fingerprinting to
test the notion that the NGC group of stars may be from Sgr.

It has long been observed
that present dSph stars are typically underabundant in [$\alpha$/Fe]
compared to Milky Way stars at the same [Fe/H], presumably a
result of the much slower enrichment history of these smaller
systems, which allows the products of Type Ia supernovae (SN Ia)
(including much of the iron) to be introduced at lower overall
metallicities. The same underabundance trend is found for various
light s-process elements, like yttrium, which are thought to be
converted to heavier s-process elements (like lanthanum) by high
neutron exposure in low mass asymptotic giant branch (AGB) stars;
thus, low Y and high La abundances are also an indication of a
slow enrichment star formation history, at least compared to the Milky Way.

Detailed abundance studies of the Sgr core have shown similar
overall trends in abundances patterns as other dSphs, but also
some trends apparently characteristic of the Sgr system itself.
For example, Smecker-Hane \& McWilliam (2002) and McWilliam \&
Smecker-Hane (2005) found that high metallicity ([Fe/H] $>$ -1)
Sgr stars show extraordinarily enhanced heavy s-process (e.g., La)
abundances, while at the same time these stars have low abundances
of Mn and Cu. Similar chemical trends are also found in the LMC
(Johnson et al. 2006; Pomp{\'e}ia et al. 2008). Since manganese
and copper yields from SN II decrease, relative to iron, in lower
metallicity supernovae, the low values of [Mn/Fe] and [Cu/Fe]
could be the result of nucleosynthesis from low-metallicity SN II,
while the large values of [La/Fe] are the product of the s-process
yields from low-metallicity, low-mass AGB stars. These patterns
can arise from intense star-formation bursts, along with loss of
some SN II ejecta via galactic winds, followed by long quiescent
periods in the dSph. After long periods of time (several Gyr),
low-mass, low-metallicity AGB stars eventually add significant
amounts of ejecta into the interstellar medium (ISM) and leave
their signature on the highest metallicity stars.

Figures 3-6 show the distributions of [Ti/Fe], [Y/Fe], [La/Fe],
and [La/Y] as a function of [Fe/H] for all of our targeted stars
(middle panels), for those of MW and other dSph stars (top
panels), and for our targeted stars superposed on those of LMC
stars (bottom panels). The MW abundance distributions have been
taken from Gratton \& Sneden (1994), Fulbright (2000), Johnson
(2002) and Reddy et al.\ (2003). The other dSph data are from
Shetrone et al.\ (2001; 2003), Sadakane et al.\ (2004) and Geisler
et al.\ (2005).  The LMC data are from Johnson et al.\ (2006),
Pomp{\'e}ia et al.\ (2008) and Mucciarelli et al.\ (2008).

It is immediately obvious from inspection of Figures 3-6 that the
bulk of our sample stars do not share the same chemical abundance
patterns as Milky Way stars. To quantify this assessment, we fit
linear trends to the Milky Way distributions (the lines shown in
Figs.\ 3-6), and determine the dispersions around those trends for
Galactic stars: 0.15 dex for [Y/Fe] and 0.12 dex for [La/Fe]. For
the [Ti/Fe] versus [Fe/H] distribution we fit two linear trends on
either side of the apparent Milky Way transition at [Fe/H]=-0.7,
and find dispersions of 0.10 and 0.05 dex to either side of that
break, respectively. (These measured dispersions represent both
the intrinsic dispersion of abundances as well as measurement
errors from the various surveys of MW stars.) For each of our M
giants we determine the number of standard deviations, $N_\sigma$,
away from the Milky Way mean trend that star is at its [Fe/H] in
each of the distributions shown ([Ti/Fe], [Y/Fe] and [La/Fe] as a
function of [Fe/H], Fig.\ 7). These deviation measures are
tabulated in columns 2-4 of Table 6. We also derive an average
deviation for all three trends (column 6 of Table 6). These
deviations allow us to characterize how ``Milky Way-like'' each
star is; we adopt as a definition of ``Milky Way-like'' those
stars that always lie within 1.5$\sigma$ of the mean MW abundance
trends. By this definition, only four stars from the 30 leading
arm north sample, no stars in the leading arm south sample, and
one star from the 13 NGC sample have abundance patterns
approximating ``MW-like''; these stars are designated with
overlying ``cross'' symbols in Figure 7 (as well as in Figure 9,
described below). For comparison, by our adopted definition of
``MW-like'' we would also classify four of 27 of the bona fide Sgr
stars from Monaco et al. and Sbordone et al. as ``MW-like''. This
simple analysis attests to the true peculiarity of the stars in
our sample by a Galactic standard, and that there is likely little
contamination by MW stars.

Note that in Paper V we had divided the Sgr leading arm north
group into a ``best'' (the fainter and farther stars that are most
likely to be in the Sgr leading arm) and a ``less certain''
subsample. The latter included stars with brighter magnitudes
($K_{s,0} < 7.5$) that we considered the most susceptible to
contamination by Galactic (thick disk) stars as well as those
stars closest to the Galactic bulge.  We showed in Paper V (see,
e.g., Fig.\ 9 of that paper) that the metallicity distributions of
the ``best'' and ``less certain'' samples were similar in shape,
spread and mean values. Further analysis of the relative abundance
patterns between these two subsamples here also reveals no obvious
distinctions: Figure 8 shows that there is no apparent difference
in the overall chemical patterns between the less certain and best
LN subsamples, even though the latter are at projected distances
more commensurate with those predicted by the Paper IV model for
the Sgr leading arm in the Northern Hemisphere and observed for
other proposed Sgr leading arm tracers (e.g., the K/M giants and
blue horizontal-branch stars (BHBs) from SDSS/SEGUE spectra by
Yanny et al. 2009, and the RR Lyrae stars (RRLSs) with large
negative Galactic Standard of Rest velocities ($v_{GSR}$) in the
Virgo stellar stream (VSS) region likely to be Sgr stars by Prior
et al. 2009b). Therefore, we continue to see no evidence that the
two LN subsamples are stars of a different origin and, given the
other evidence presented here that they are all most likely to be
Sgr stream stars, we continue to consider them together as the
single ``Sgr leading arm north sample" in this paper.\footnote{It
is perhaps worth mentioning that present Sgr stream models, such
as that presented in Paper IV, show that the leading arm may wrap
a second time around the Galactic center --- and with the debris
in the second leading arm wrap lying closer to the Sun in the
Northern Hemisphere than the debris in the first wrap. This is a
potential source of contamination of the LN sample by further
extensions of itself; however, if the trend between the LN and LS
stars is extrapolated, stars in a second leading arm debris wrap
should be even more metal-poor on average than the LS sample. That
there is no clear distinction in metallicities between the closer
and farther LN stars --- while the LS stars {\it are} more metal
poor on average --- suggests that this ``self-contamination'' of
the LN sample by debris from the second wrap of the Sgr leading
arm may not be significant.}

Figure 9, which shows the distribution of [Y/Fe] versus [Ti/Fe],
further demonstrates the true distinction between our sample stars
and those in the standard Galactic populations; stars in our Sgr
sample are shown color-coded by their Sgr system grouping in the
top panel. We also compare the MW stars with those of other dSph
and LMC stars on the middle and bottom panels, respectively.
Figure 9 shows a striking segregation of stars by their parent
system (particularly by [Y/Fe]), and one that further illustrates
the differences in chemical evolution between stars we have
selected to be Sgr-members and the nominal MW populations (see
\S5). In the top panel of Figure 9 we have marked with large
crosses those stars that were deemed most MW-like by the above
$\sigma$ analysis; as might be expected, these tend to lie closest
to the V-shaped distribution of MW stars, but even then not in the
main locus of MW stars, but, rather, skirting it. Figure 9 further
reinforces the conclusion that the sample of M giants we have
observed should have very little contamination by nominal MW field
stars, and this includes stars we have selected to lie in the NGC
group (see \S4-2). Furthermore, we notice that stars from the
other dSphs as well as the LMC sit in the bottom left part of
Figure 9 (middle and bottom panels), so that they share a similar
--- but not exactly the same
--- distribution as our Sgr stars. It is worth noting that about a
third of the stars selected to be from the Sgr system (including
one or two Sgr core stars) have higher [Ti/Fe] than the bulk of
the dSph and LMC stars (a feature seen also in Fig.\ 3).
Nevertheless, since about 2/3 of the ``Sgr system'' stars do share
the same chemical patterns as other dwarf galaxy systems in Figure 9, we can not rule
out the possibility that a small number of these particular Sgr
stars are from other dwarf satellite debris. We discuss this point further in
\S4.3.

The observed deviation of our selected Sgr stream stars away from
MW abundance trends as a function of [Fe/H] is illuminating.
Deviation trends can be used as a quantitative chemical marker,
and combinations of abundance patterns for different elements are
thought to provide unique, or at least fairly distinctive,
chemical fingerprints that can be used to identify the star
formation sites (e.g., the parent galaxies) of specific stars. For
example, inspection of the [Ti/Fe] vs [Fe/H] trends shown in
Figure 3 (top and middle panels) and Figure 7 reveal clear
differences between our sample of likely Sgr stars and MW stars:
While there is a general agreement in the [Ti/Fe] levels for the
MW and Sgr stars at the lowest metallicities ([Fe/H]
$\lesssim$-1.0) --- where both the Sgr and MW stars are equally
enhanced in [$\alpha$/Fe], a feature that betrays the signature of
prevalent SN II enrichment --- the differences in abundance trends
increase rapidly above [Fe/H] $\sim -1.0$, with the great majority
of Sgr stream stars falling below the MW trend. The latter trend
reflects the lower star formation efficiency and/or slower
chemical enrichment (and, thus, greater relative SN Ia yields) of
Sgr stars relative to those of the MW (see \S5).

Meanwhile, the trend for [Y/Fe] shows the Sgr stars to be
underabundant with respect to the MW and subsolar over the whole
range of sampled [Fe/H] (Figs. 4 and 7).  On the other hand,
[La/Fe] is primarily below the MW trend until [Fe/H] $\sim -0.5$,
when it quickly rises well above the MW level. This trend was
previously found by Smecker-Hane \& McWilliam (2002). The relative
abundances of heavy to light s-process elements shown by [La/Y]
(Fig. 6) accentuate the differences between Sgr and MW stars at
all metallicities.  We further interpret the meaning of these
various trends in \S5.

\subsection{Chemical Trends Along the Sgr Stream and Chemical Fingerprinting of the \\
North Galactic Cap Stellar Group}

Presently, the only known dSph satellite of the MW known to
contain a significant population of M giants is the Sgr system.
While the Magellanic Clouds contain large numbers of M giants, the
analyses undertaken in Paper I and elsewhere have revealed no
evidence of any M giant tidal structures from either of the
Magellanic Clouds. At low Galactic latitudes M giants are found
associated with the Monoceros/Galactic Anticenter Stellar
Structure (GASS, Rocha-Pinto et al. 2003), but it is still not
clear whether this structure is a tidal stream or a part of the MW
disk (e.g., Martin et al. 2004b; Momany et al. 2004, 2006;
L\'{o}pez-Corredoira et al. 2007). The only other MW substructure
that is certainly a tidal debris remnant and known to contain M
giant stars is the more distant, Triangulum-Andromeda (TriAnd)
star cloud (Rocha-Pinto et al. 2004), also at low Galactic
latitudes ($-40 ^{\circ}< b < -20 ^{\circ}$). For higher Galactic
latitudes, the analysis of M giant distributions in Paper I showed
that a major fraction ($\gtrsim 75\%$) of M giants found in the
halo away from the Galactic plane lie along the Sgr orbital plane;
since overall the presence of M giants in other MW dSphs or
substructures is relatively rare, the concentration of high
latitude Galactic M giants along the Sgr plane suggests that the
vast majority of these stars were indeed contributed by the Sgr
dSph.

This proposition received further support with subsequent study of
the radial velocities of these M giants, which show most of them
also to lie in correlated trends of velocity with Sgr orbital
plane longitude (Majewski et al. 2004, ``Paper II'' hereafter;
Paper IV) corresponding to the Doppler motion of the most recently
stripped (e.g., within the last 1-2 Gyr) Sgr debris as it wraps
around the Galactic center (see Paper IV). Recently, Yanny et al.
(2009) have shown that our adopted radial velocity (RV) sequence
for Sgr stars in the northern sky (e.g., as shown in Figure 2 of
Paper V) is well matched by SDSS/SEGUE-derived radial velocities
for candidate Sgr K/M giants and blue horizontal-branch stars
(BHBs) along the leading arm. This correspondence lends further
credence to the notion that most of our RV-selected Sgr leading
arm stars are indeed from the Sgr dSph.

However, not {\it all} M giants in the Sgr plane are found to lie
along the primary velocity trends of this most recently lost
debris. In particular, a number of radial velocity ``outliers''
have been found in the southern Galactic hemisphere, mostly in the
Sgr longitudinal range $\Lambda_{\odot} = 20-90^{\circ}$ (Paper
II). These particular stars have been associated with older parts
of the Sgr leading arm that have already passed below the Galactic
plane and that form a new loop around the Galactic plane (Paper
IV); a fraction of these stars constitute our ``Leading Arm South
(LS)'' sample in Paper V, where we show them to have a more
metal-poor distribution overall than the dynamically younger,
``Leading Arm North (LN)'' sample. Figures 3-7 in this paper show
these LS stars to also have abundance patterns indicating that
they are not only dynamically older Sgr stream stars (i.e.
stripped from Sgr longer ago), but also likely to have actually
formed earlier than the LN stars on average. Specifically, the LS
stars show a higher mean [Ti/Fe] level (0.26 dex) than that of
either the Sgr core ($-0.05$ dex) or LN (0.13 dex) samples, which
demonstrates that, on average, the LS stars were formed when
enrichment in Sgr was still dominated by SN II. This difference in
[Ti/Fe] trends between the LS and LN is consistent with the
age-metallicity relation of Sgr (Siegel et al. 2007), which shows
that stars of the mean metallicity of the LS ([Fe/H] $=-1.1$) were
formed on average $\sim$11 Gyr ago, only $\sim$1-2 Gyr after the
oldest known populations in Sgr and before significant numbers of
SN Ia progenitors (mass transfer binaries) from those earliest
populations could have evolved to supernova stage and introduced
significant iron yields into the Sgr ISM. On the other hand,
Siegel et al.'s age-metallicity relation suggests that stars of
the mean metallicity of the LN, [Fe/H] $\sim -0.7$, were formed
about 7.5 Gyr ago, or well after the onset of routine SN Ia
enrichment.  As pointed out in Paper V, that the dynamical age
difference between the LN and LS stars is only about 0.8 Gyr
whereas the mean metallicity difference between these samples is
$\Delta$[Fe/H] $\sim -0.4$ dex (and that both of these differ
significantly in mean metallicity from that of the core stars, at
$<$[Fe/H]$>=-0.4$) suggests that the Sgr progenitor must have had
a significant radial metallicity gradient before disruption.  The
age-metallicity relation of Siegel et al. implies that there must
have been a significant mean age gradient as well.

A second group of M giant stars that do not lie along the primary
velocity sequences of more recently stripped Sgr stream stars are
those we have called the North Galactic Cap (NGC) group, which
have velocities more positive than the LN stars in the same part
of the sky (Paper II). In Paper V we showed that our sample of
these stars has a similar metallicity distribution to that of the
LS sample, and, by reference to the Paper IV model, we concluded
by their position, velocity and metallicity distribution that the
NGC group might be the trailing arm counterpart above the Galactic
plane to the Leading Arm South group below the Galactic plane ---
that is, the two groups are Sgr debris of the same dynamical age.
Figures 3-6 and Figure 9 further support this notion by showing
that the abundance patterns of the NGC group very closely align
with those of the LS.  Not only does this comparison chemically
fingerprint the NGC stars as likely Sgr debris, but it places
these stars into the Sgr disruption dynamical time sequence, and
in a place in that sequence that makes sense within the context of
the Paper IV model for trailing arm debris in the North Galactic
Hemisphere that was stripped at the same time as the LS stars,
i.e. roughly 3 Gyr ago (see green debris in Fig.\ 1 of Paper IV or
Fig.\ 1 of Paper V).  By tagging the NGC group of stars to the Sgr
trailing arm through its abundance patterns we demonstrate here
one of the earliest direct applications of the concept of
``chemical fingerprinting''.

\subsection{Caveats and Alternative Scenarios}

Despite the above conclusions regarding the association of all
four of our target subsamples to the Sgr system, the recent
discovery of the ``Virgo stellar stream'' (VSS, Duffau et al.
2006; Vivas et al. 2008) and ``Virgo overdensity'' (VOD, Juri{\'c}
et al. 2008) shows these features to span a wide range of Northern
Galactic Hemisphere sky (over 1000 deg$^{2}$ in the case of the
VOD; Juri{\'c} et al. 2008), and it is worth evaluating whether
any of our subsamples --- namely the Sgr LN stars and the NGC
groups, which are also in the Northern Galactic Hemisphere ---
could be related to these extensive Virgo structures. Both the VSS
and VOD have been explained in terms of debris from dSph mergers
with the MW, so could account for high northern latitude stars
with non-MW-like abundance patterns. We compare five properties of
the LN and NGC groups against those of the VSS and VOD to
demonstrate that there is unlikely a connection between the latter
two halo substructures and the LN and NGC group stars.

(1) {\it Distances:} The distance of the VSS is $\sim19$ kpc
(Duffau et al. 2006; Newberg et al. 2007; Prior et al. 2009),
while the VOD is estimated to have a distance of $\sim$6-20 kpc
(Juri{\'c} et al. 2008; Vivas et al. 2008; Keller et al. 2009).
Using the [Fe/H] values measured for the M giants in Paper V, we
have used the corresponding isochrones from Marigo et al. (2008)
and the observed, dereddened $K_s$ magnitudes to estimate
photometric parallax distances to each of our stars. This exercise
is one fraught with large uncertainties, since for each star we
have to assume an age (which can be roughly deduced from the
age-metallicity relation, e.g., that in Siegel et al. 2007), a
mean [$\alpha$/Fe] (which may or may not be well traced by
[Ti/Fe]), and whether a particular star is on the first or second
ascent giant branches; nevertheless we attempted the exercise to
get a rough idea of whether the stars could be at the distances of
the VSS or VOD.

Based on the rough photometric parallaxes we find that all of the
LN and NGC stars are closer than the VSS, whereas a large number
of the LN and NGC stars are in the range of distance quoted above
for the VOD. Fortunately, other properties of these stellar
systems give more definitive discrimination between them (see
below).

(2) {\it Sky Positions:} The VSS is a narrow structure currently
mapped from $l = 279^{\circ}$ to $317^{\circ}$ and $b =
60^{\circ}$ to $63^{\circ}$ (Duffau et al. 2006); even
extrapolating that swath along the same great circle, the VSS
hardly intersects the region of the sky covered by the NGC group
sample, which, in any case, is much more broadly dispersed than
the relatively narrowly confined, $\sim 3^{\circ}$-wide VSS. The
positions of only two LN stars overlap the extrapolated swath of
the VSS, but the velocities of these two stars (stars
1236549$-$002941 and 1319368$-$000817) grossly mismatch that
expected for the VSS at their positions.

On the other hand, the VOD {\it is} much more broadly distributed
on the sky, covering around 1000 deg$^2$ with a center at
$(l,b)\sim (300^{\circ},65^{\circ})$; however, the area of the sky
covered by this excess (Juri{\'c} et al. 2008) is very different
than the areas covered by either the LN or NGC samples. Only a
mere seven stars from the LN and one star from the NGC group
subsamples even lie in the most liberal definition of the angular
extent of the VOD.

(3) {\it Velocities:} A number of VSS stars have had spectra taken
that show them to have $v_{GSR}$ spanning $\sim100-130$ ${\rm
km\,s^{-1}}$ (Duffau et al. 2006; Newberg et al. 2007; Prior et
al. 2009).  The only available velocity information on the VOD
comes from the QUEST RR Lyrae star survey (Vivas et al. 2008)
combined with a sample of BHBs from the SDSS survey (Sirko et al.
2004). Vivas et al. find that there are three moving groups in the
VOD region, with average $v_{GSR}=+215$, $-49$ and $-171 {\rm
km\,s^{-1}}$, respectively. These authors suggest that the VOD
actually consists of the VSS plus other halo substructures. The
stated $v_{GSR}$ range of the VSS has the opposite sign as most of
the Sgr LN stars, but does cross the range of $v_{GSR}$ spanned by
our NGC group M giants. However, no more than three of our NGC
sample stars, and five of the LN sample stars barely match {\it
both} the position and velocity distribution of either the VSS or
VOD.

(4) {\it Metallicities:} The M giants in our Sgr sample stretch
broadly from solar metallicity down to the the lowest metallicity
where M spectral type giants can form ([Fe/H]$\sim-1.4$), but
there is no correspondingly high metallicity population that has
been identified with the VSS and VOD thus far. The metallicity of
VSS RR Lyrae stars is $\sim -1.86$ to $-1.95$ (Duffau et al. 2006;
Prior et al. 2009), and VOD main sequence stars are at $\sim -2.0$
(An et al. 2009). Since the metallicity of RR Lyraes can extend to
solar metallicity (as seen, for example, in the local MW disk;
Layden 1994), the lack of metal-rich RRLSs found in the VOS or VVS
suggests that the stellar populations in the VSS/VOD are
metal-poor in the whole, and even slightly more metal-poor than
the RR Lyrae stars  found among the Sgr tidal debris (which have
[Fe/H]$\sim-1.7$ to $-1.8$; Vivas et al. 2005, Prior et al. 2009b,
Starkenburg et al. 2009).

(5) {\it Evidence for M giant populations:} Of course, the very
existence of M giants in a stellar population is determined by its
metallicity, with only relatively metal-rich populations making
first ascent giant branch stars that late in spectral type. There
is no evidence for M giant stars distributed kinematically and
positionally as counterparts to the other VSS/VOD tracers. If
either the VSS or VOD structures contained an M giant population,
one might expect at least a concentration of those M giants at the
densest, ``core'' parts of those structures as traced by other
stellar types; yet the analysis of 2MASS M giants in Paper I, as
well as more recently by Sharma et al. (2009, in prep.), reveals
no concentration of M giants either centered on the VOD or
following the VSS. On the other hand, that the Sgr stream {\it
does} contain M giants, that the M giants in the Galactic halo are
almost entirely concentrated along the Sgr orbital plane (except
at the very lowest latitudes, where Monoceros/GASS and TriAnd
contribute), and that the M giants we have selected are generally
matched in position on the sky with known Sgr features makes it
far more likely that the LN and NGC group stars are associated
with Sgr than the VSS or VOD.

On the other hand, there does remain an alternative scenario to
that we have adopted as most likely here (namely that all of our
subsamples are related to the Sgr system) that is presently very
difficult to discriminate against. It has recently been proposed
that all or some satellites of the MW may have been accreted as
one or more groups of galaxies (Li \& Helmi 2008; D'Onghia \& Lake
2008; Metz et al. 2009). For example, D'Onghia \& Lake (2008)
proposed that an ``LMC group'' --- composed of the LMC, Small
Magellanic Cloud (SMC), Sgr and other satellite galaxies --- were
originally part of a collection of galaxies once bound to each
other and that later fell into the MW together. This hypothesis
can explain the planer orbital configuration of some dSphs in the
MW halo (e.g., Kunkel 1979; Lyden-Bell 1982; Majewski 1994; Palma
et al. 2002; Metz et al. 2008), and could explain the planar
configurations of tidal debris, even if that debris ultimately
derived from different satellites. With slight differences in
initial orbits of the parent dwarf galaxies, that debris today
might have differing orientations and velocities, despite being in
nearly the same plane and having the same general direction of
angular momentum; in principle, therefore, one could explain our
various M giant samples lying in a largely planar distirbution as
deriving from different parent accreted dwarf galaxies, but
perhaps parents that fell into the MW together.

One motivation for proposing such an ``infalling group'' scenario
in the specific case of the Sgr system is that it might provide an
origin for the recently discovered, but not yet
definitively-explained bifurcation of the northern Sgr stream
leading arm as seen in the SDSS imaging (Belokurov et al. 2006;
Fellhauer et al. 2006).\footnote{We point out that Fellhauer et
al. (2006) attribute one each of the two Sgr structures in the
SDSS Northern Galactic Hemisphere data to the leading and trailing
arms of the Sgr stream, respectively. But this explanation is not
consistent with the predictions of the distances of these
structures by the Paper IV model, which was designed to constrain
not only the positions of Sgr data on the sky (the main criteria
used by the Fellhauer et al. 2006 Sgr model), but the extant Sgr
stream velocity data as well. Furthermore, Yanny et al. (2009)
suggest that the two branches might be from debris that was
stripped at similar times, due to the similar velocities,
metallicities, and relative densities of K/M giant, BHB, and
F-turnoff stars.} On the other hand, Yanny et al. (2009) find not
only the metallicities, but the velocities and distances to be
indistinguishable in the two pieces of the bifurcated SDSS Sgr
feature. This suggests a strong coherence in star formation and
chemical enrichment histories, at least for those stars along the
bifurcated, Northern Hemisphere Sgr arm --- and, were such a
coherence between two parent systems to exist, it would be very
hard at present to distinguish this from a single parent origin
(i.e. the known Sgr core) for all of the debris. At present, in
the absence of evidence supporting multiple parent systems
producing multiple M giant substructures along the same plane in
the sky, we prefer the most straightforward interpretation for the
connection of the LN and LS groups as parts of a single Sgr
leading arm, as well as the identification of the NGC group with
(a diffuse distribution of) the Sgr trailing arm.

In conclusion, detailed chemical abundance analysis of our various
samples demonstrates that these stars are by and large dSph-like
and with little contamination by the nominal Milky Way
populations, halo or disk. Because the leading arm stars were also
pre-selected to be in the Sgr stream and to follow the expected
velocity trends for Sgr debris, because they form a clear and
logical chemical sequence, and because no evidence for other M
giant tidal debris from any other satellite is found to intersect
the Sgr stream in relevant parts of the sky, and, furthermore,
because all evidence continues to support that the bulk of all
high latitude M giants have been contributed from the Sgr system,
we conclude that the vast majority of our leading arm stars must
be from the Sgr dSph. Since the NGC sample stars share similar
chemical patterns of Sgr LS stars, we conclude that they are
likely Sgr stars as well, but this hypothesis requires them to be
from the trailing arm debris due to their peculiar velocities.
Therefore, we propose that all of our M giant samples, taken as a
group, together paint a chemical portrait of the Sgr dwarf as it
appeared $\sim3$ Gyr ago.

Finally, we close this discussion of caveats with a warning to the
reader that although we believe the LN, LS and even the NGC group
stars to be primarily constituted by stars stripped from the Sgr
dSph system, these samples represent highly biased sets of such
stars by distance: In the present high resolution spectroscopic
study we selected targets that we expected to be Sgr stream
members but that were also bright enough to allow echelle
spectroscopy to adequate $S/N$ for chemical abundance analysis.
Moreover, we selected the target stars for this study (and that in
Paper V) from an input catalog
--- those M giants that had radial velocities from medium resolution spectroscopy
(e.g., Paper II, Paper IV) --- that itself shares some degree of
the same bias. Given this strong prejudice for the closest,
brightest Sgr members we could find, one should exercise due
caution in how these stars are used. For example, one should not
use the stars in Table 3 to constrain the mean distances of
different parts of the Sgr stream since those mean distances no
doubt lie beyond the stars presented here; the latter are likely
several or more $\sigma$ outliers in the distance distribution of
Sgr stream stars.

\section{Global Chemical Evolution of the Sgr System}

We can take advantage of our entire Sgr high resolution core +
tidal tail sample to build the most complete (in terms of
metallicity span) chemical portrait of the Sgr progenitor yet
assembled; this portrait is summarized by the middle panels of
Figures 3-6 as well as Figure 9. Chemical evolution is driven by
the cycle of nucleosynthesis and subsequent transfer of
nucleosynthetic products into the environment for incorporation
into future generations of stars. Chemical evolution models guide
the interpretation of observed patterns in a galactic system such
as those demonstrated in Figures 3-6 and 9.

\subsection{[Ti/Fe]}

For example, as mentioned above, $\alpha$ elements are mainly
produced by SN II while iron is synthesized largely by SN Ia,
whose progenitors have a lifetime of $\sim1$ Gyr. Therefore,
[$\alpha$/Fe] is high for early chemical enrichment in a stellar
system and then declines as SN Ia ``turn-on". Titanium acts mainly
as an $\alpha$ element, and from its trend with [Fe/H] we can
infer how far chemical evolution proceeded in the first $\sim1$
Gyr of the system's life, an indication of the initial star
formation rate (SFR). Figure 3 shows that the downturn in [Ti/Fe]
happens at around [Fe/H] $=-0.9$ for Sgr, which is only a few
tenths of a dex lower than the transition seen in the MW field
star population; this suggests a lower early SFR in Sgr than for
the MW field population (Monaco et al. 2005b; Sbordone et al.
2007). Alternatively, Lanfranchi et al. (2006) argue that
initially Sgr may have started out with a high star formation
efficiency --- at least higher than that of other dSphs --- but
the resulting intense galactic wind acted to substantially squelch
the SFR. Lanfranchi et al. claim that most of the observed Sgr
stars formed after the beginning of the wind, which explains why
they have lower [$\alpha$/Fe] than MW stars.

Figure 3 also shows the [Ti/Fe] trends for the LMC and other
dSphs, which are very different from the trends for the MW and the
metal-poor Sgr stars. The chemical differences among these various
systems are due to their unique star formation histories (Venn et
al. 2004; Johnson et al. 2006). As may be seen, in the LMC and
other dSphs, the [Ti/Fe] is low over all [Fe/H] probed, due to a
lower early SFR, which is the result of low star formation
efficiencies and high galactic winds (Geisler et al.\ 2007), and
also probably because of fewer high-mass SN II in these small
systems (Woosley \& Weaver 1995; Tolstoy et al.\ 2003; Pomp{\'e}ia
et al.\ 2008).

We also find the metal-poor Sgr stars (with [Fe/H] $\lesssim-1.2$)
to have high Ti abundances, similar to the trend for the MW. This
implies that the early chemical composition of Sgr was more like
the MW (Shetrone 2004) than the LMC and other dSphs; from a
similar comparison of [$\alpha$/Fe] abundances of low [Fe/H],
Monaco et al.\ (2005b) conclude that the Sgr progenitor was
probably a relatively large, star forming, gas-rich object. On the
other hand, Monaco et al. (2005b) also suggest Sgr should have had
a different subsequent chemical evolution from the MW and other
Local Group galaxies due to the strong and disruptive dynamical
interactions this system has clearly had with the MW; these
interactions in a gas-rich system can typically trigger star
formation activity (e.g., Kravtsov et al.\ 2004; Zaritsky \&
Harris 2004).

\subsection{s-process Elements}

The s-process elements are thought primarily to occur during
thermal pulses in the intershell convection zone in low mass AGB
stars. The neutron flux per seed nucleus is roughly inversely
proportional to the metallicity of the AGB.  Therefore, at low
metallicities AGB stars produce heavier s-process elements like La
more efficiently than lighter species like Y, all of the way up to
the formation of the heaviest s-process element, Pb, in the most
metal-poor environments (see the review by Busso et al.\ 1999 and
2004). Figures 5-6 show upturns in [La/Fe] and [La/Y] for Sgr at
[Fe/H] $\sim-0.5$.  Since [La/Y] is enhanced in metal-rich Sgr
stars, the high ratio of heavy to light s-process elements
([hs/ls]) among the metal-rich stars indicates a strong
contribution from low-metallicity AGB progenitors (Smecker-Hane \&
McWilliam 2002), and a slower SFR of Sgr than the MW, so that the
low-metallicity AGB yields have enough time to contaminate the ISM
(Venn et al.\ 2004; Pomp{\'e}ia et al.\ 2008).  On the other hand,
Sgr and LMC stars at lower metallicity show similar trends to MW
stars, particularly in Y and La; this indicates that these stars
might be among the first ones formed because they were not yet
contaminated by the AGB yields (Johnson et al. 2006; Mucciarelli
et al. 2008). Moreover, models for the evolution of old, gas-poor
dSphs (e.g., Sculptor) predict subsolar Y over all [Fe/H] probed,
and an upturn in La abundance at [Fe/H] $\sim-1.6$, due to the
appearance of the products of metal-poor AGB progenitors in the
ISM (Fenner et al.\ 2006; Gibson 2007). The patterns of La in Sgr
basically agree with the prediction of the gas-poor dSphs, except
the upturn in Sgr occurs at higher [Fe/H].  That suggests a higher
early SFR of Sgr than other dSphs, and the high [hs/ls] can be
seen as another ``clock'' of star formation, in addition to the
[$\alpha$/Fe] ratio.

\subsection{Relative Galaxy Star Formation Rates}

Figures 3-6 have demonstrated abundance trends for Sgr that
distinguish it from the MW and other MW satellites. Several of
these differences relate to the onset of specific chemical
chronometers, such as the contribution to the gaseous environment
of the yields from SN Ia (seen as a decrease in [Ti/Fe]) and
metal-poor AGB (seen as an increase in [La/Y], for example), where
the [Fe/H] corresponding to the chemical signature is correlated
to the SFR prior to the creation of that signature. A comparison
of the above two chronometers ([Ti/Fe] and [La/Y]) suggests
a natural sequence in relative SFRs from the other MW dSph systems
(lowest SFRs), to the LMC (a modest SFR), and Sgr (the highest SFR
among the MW satellites), based solely on the positions of the
transitions in chemical properties in [Fe/H].

To emphasize the point that the chemical histories of the MW
satellites likely differ primarily as a function of their SFRs, we
modify the comparisons of the chemical properties of Sgr to those
of the LMC and other dSphs as shown in Figures 3-6 by shifting the
distributions of the dSph ensemble and the LMC by $\Delta$[Fe/H] =
+1 and +0.4, respectively (these particular values were selected
by eye for illustrative purposes only). The results, shown in
Figure 10, reveal a much closer agreement in the overall shapes of
the abundance trends shown, suggesting even the possibility of a
``universal'' chemical enrichment pattern among MW satellites,
with the primary difference being the [Fe/H] placement of the
pattern, which itself is a function of the early SFR.

It is immediately obvious from Figures 3-6 and Figure 10 that Sgr
much more closely matches the overall chemical evolution of the
LMC than it does other MW dSphs. Indeed, their similar chemistries
suggest that it may be reasonable to consider the LMC --- a late
type, dwarf spiral or irregular galaxy --- as a more appropriate
paradigm for the pre-interaction state of Sgr than is provided by
the other dSphs, even though Sgr morphologically resembles the
other dSphs now in terms of its more regular structure and lack of
current star formation or gas.   Sgr's present morphological
difference with the LMC is easily accounted for by the fact that
the LMC, which apparently has just fallen into the MW environment
for the first time (Kallivayalil et al.\ 2006; Besla et al.\ 2007;
Piatek et al.\ 2008), has not experienced the tidal battering that
the MW-bound Sgr has experienced for at least the past few Gyr. It
is well known that severe tidal encounters such as has been
experienced by Sgr can ``tidally stir'' dwarf irregular systems
into dSphs (Mayer et al.\ 2001; Skillman et al.\ 2003;
Klimentowski et al.\ 2007). Evidence that as recently as several
Gyr ago Sgr was actively forming stars can be found in its
color-magnitude diagram, which shows evidence for populations as
young as 2 Gyr old or even younger (Sarajedini \& Layden 1995;
Layden \& Sarajedini 2000; Siegel et al. 2007).

Typically SFRs and chemical enrichment are thought to be driven by
the mass of a system. The mass-metallicity relation of dSphs has
been investigated by, for example, Yoshii \& Arimoto (1987) and
Tamura et al. (2001), who find that metallicity is roughly
logarithmically proportional to the mass of these satellite
galaxies. Because Sgr is presently dominated by stellar
populations with higher metallicity, that it is more massive than
the other MW dSphs is consistent with the notion of a
mass-metallicity relation. Curiously, however, Sgr is more
chemically evolved than the LMC, despite the fact that the mass of
Sgr, at least that estimated the system had several Gyr ago by
Paper IV of $\sim 2-5 \times 10^8$ M$_{\odot}$, is less than that
of the LMC ($\sim 1-2 \times 10^{10}$ M$_{\odot}$; Gardiner \&
Noguchi 1996; van der Marel et al.\ 2002). Certainly star
formation efficiency may also be a significant driver in the rate
of chemical evolution (Lanfranchi et al.\ 2006, 2007), but an
expected consistency with mass-metallicity relations, and the
possibility of a universal enrichment pattern (e.g., Fig.\ 10),
may hint that Sgr was actually once significantly more massive
than even the more recent, high estimates of its former mass,
perhaps with a mass comparable to or larger than the LMC.  If so,
this would imply a much longer tidal stripping history for Sgr
than has been previously observed (e.g., Paper I; Belokurov et
al.\ 2006) and modeled (Paper IV; Fellhauer et al.\ 2006), perhaps
with much of that mass lost as pure dark matter (to explain the
lack of obvious stars from earlier orbits than has heretofore been
observed).

\section{Summary}

Chemical abundances in stars are fossil records of the enrichment
history of a galaxy and their distinctive patterns provide
signatures that, if not uniquely branding its stars, at least
allow us a means to test whether particular stars are likely to be
associated with that system based on whether their chemistry fits
into its overall abundance patterns. We have applied this test
here not only to demonstrate the likely high purity of the
spatially- and kinematically-selected sample of M giant Sgr tidal
tail star candidates used in Paper V to show the existence of a
strong metallicity gradient along the Sgr tails, but also to prove
the likely Sgr-origin of the somewhat mysterious ``North Galactic
Cap moving group'' also discussed in that previous contribution.
We conclude that most of the NGC moving group stars are probably
from the old trailing debris arm of Sgr (their MDF also supports
this conclusion, see Paper V). Thus the present paper demonstrates
the applicability of ``chemical fingerprinting'', a technique long
discussed as one of the potentially valuable future tools of
stellar populations research.

The evolving chemical content of a galaxy depends on many
variables --- such as the stellar initial mass function and the
SFR --- that lead to the specific distribution of chemical
patterns among the stars in the system. Tidal stripping can also
shape the {\it observed} abundance patterns of the stars in a
galaxy by preferentially removing certain populations,
particularly if the system has spatial variations in metallicity
and/or an age-metallicity relation over timescales overlapping the
period during which the tidal loss of stars occurs. Both of these
situations are at play in the Sgr system. Only by surveying both
the stars lost from as well as remaining in a galaxy like Sgr do
we have hope of recovering an unbiased view of its chemical
history. We have explored the chemical abundance patterns in stars
along the Sgr tidal tails, which, when combined with data on stars
in the Sgr core, afford us the most complete view of the original
chemical abundance distributions of this system to date; but even
then, our view here is likely to be rather incomplete and still
tells us only part of the story. For example, we still do not have
good information on the chemical properties of the oldest, most
metal-poor Sgr populations.

In stars associated with Sgr we have found lower values of [Ti/Fe]
at [Fe/H] $> -1$ than for MW stars at the same [Fe/H] (Fig.\ 3).
And while the MW exhibits an apparent transition of [Ti/Fe] to
solar-like levels at [Fe/H] $\sim -0.7$, such a transition (and to
even lower [Ti/Fe] levels) happens for Sgr stars at [Fe/H] $\sim
-1$. This shows that Sgr had a slower SFR than the MW, like other
Galactic satellites. We also find the Sgr stars to have subsolar
abundances of the light s-process element yttrium at all [Fe/H]
probed, while the heavy s-process element lanthanum is enhanced
relative to the MW for [Fe/H] $\gtrsim -0.5$ (Figs.\ 4-6). This
indicates the importance of low-metallicity AGB nucleosynthesis in
the metal-rich Sgr stars (Smecker-Hane \& McWilliam 2002) and is
another signature of a slower SFR than occurred in the MW.

We also find that although the LMC and other dSphs exhibit similar
chemical pattern trends as Sgr, these patterns exhibit their
significant abundance transitions at lower [Fe/H] than for Sgr.
Such differences suggest a faster enrichment and more rapid star
formation history in Sgr relative to those in the LMC and other
dSphs. After applying a hypothetical shift in [Fe/H] of +1 dex in
the abundance patterns of other dSphs and +0.4 dex for the LMC we
find that the chemical patterns of Sgr, LMC and other dSphs
strongly resemble each other (Fig.\ 10), which suggests the
possibility of a universal chemical enrichment progression among
MW satellites that differs only by the SFRs; these SFRs, in turn,
are likely correlated to the original masses of these systems. The
relative shifts suggest that the relative SFR of these galaxies
are, in order from slowest to fastest: the other MW dSphs, the
LMC, Sgr and the MW respectively.  Based on their close chemical
similarities we suggest that Sgr was probably formerly more
similar in nature to the present morphology and structure of the
LMC than to those of other MW dSphs. The wide ranges of
metallicities in Sgr suggests a large age spread, and implies a
long duration of star formation in its central regions
(Smecker-Hane \& McWilliam 2002).  This agrees with the discovery
of quite young populations in the Sgr core (Siegel et al.\ 2007),
which shows that Sgr has had even relatively recent star formation
activity, like the LMC.

The color-magnitude distribution of Sgr stars demonstrates that
its star formation history was highly variable, including some
fairly well-defined ``bursts'' (Layden \& Sarajedini 2000;
Bellazzini et al.\ 1999; Monaco et al.\ 2002; Siegel et al.\
2007). These produced populations with different, but overlapping
radial density profiles in the progenitor satellite, and likely a
strong internal metallicity gradient. In Paper V we found a strong
metallicity gradient along the Sgr tidal arms, with the stars in
these arms increasingly more metal-poor with angular separation
(i.e. orbital phase difference) from the core. By comparing this
gradient to models of the timescale for the disruption that
produced these tails (Paper IV) we argued in Paper V that Sgr must
have experienced a quite rapid change in its binding energy over
the past several Gyr, even if the Sgr progenitor had one of the
steepest metallicity gradients among the known MW dSphs. These
multiple population bursts also created Sgr's unique chemical
patterns, especially at higher metallicities, compared to those of
the MW (as well as to those of other dSphs). Nevertheless, while
the abundance patterns of stars presently in the Sgr core differ
greatly from those of MW stars of the same metallicity, in this
paper we have found that the more metal-poor stars in the Sgr tail
actually have abundance patterns that more closely resemble those
of MW stars at their respective [Fe/H]. The Sgr example vividly
demonstrates that while the current populations of stars in dSph
satellites are indeed chemically differentiated from the MW field
population and that ``one could not build the MW halo from the
{\it present} MW satellites'' (as emphasized by, e.g., Unavane et
al. 1996), this point is not very relevant because it is still
possible that the MW halo field population could have derived from
the {\it stripped off} populations of these very same satellites.
Majewski et al. (2002) and Mu\~noz et al. (2006) have previously
made the same point using the Carina dSph example.

The Sgr system can be seen as an evolutionary bridge from dSphs to
the MW in galaxy evolution. On the one hand, recent models
(Robertson et al.\ 2005; Font et al.\ 2006a,b) predict that the
local halo assembled rapidly --- before SN Ia had time to occur
--- with the early accretion and dissipation of a few massive
satellites to produce the high values of [$\alpha$/Fe] at low
[Fe/H] seen among halo field stars. Font et al.\ suggest the MW
satellites accreted $\sim$9 Gyr ago have all been disrupted
completely, while the surviving satelliites of today were only
recently accreted into the MW system within the past few Gyr on
relatively circular orbits.  The implication is that these
satellite systems have yet to contribute stars to the halo. On the
other hand, Sgr clearly provides an example of an ongoing merger
event that is not only contributing stars to the halo, but some
low metallicity stars of high [$\alpha$/Fe]; it may therefore be
seen as a counterexample to the ''early accretion'' hypothesis.
However, Font et al.\ point out that none of the present MW
satellites are presently located in the inner halo except Sgr and
because of this Sgr may be an exceptional case. But the recent
work on the Ursa Minor (Mu\~noz et al. 2005), Leo I (Sohn et al.
2007; Mateo et al. 2008) and Carina dSphs (Mu\~noz et al. 2006,
2008) yielding evidence for tidally stripped stars from these
systems suggests that the present satellites may also have already
contributed stars to the MW halo. This, coupled with the finding
that some of the MW dSph stars at the lowest metallicities are
indeed $\alpha$-enhanced (Shetrone et al. 2003; Geisler et al.
2005; Tolstoy 2005), suggests that Sgr may not be the only current
contributor of such stars; at the very least these systems are
likely to be {\it future} contributors of these stars, which shows
that they did not all originate in early accretions.

The results of this paper would be considerably strengthened by a
careful survey of the chemical abundance patterns of Sgr trailing
arm stars. The significant overlap in orbital phase position along
the Sgr leading arm (see Fig.\ 1 in Paper IV) ``fuzzes out'' the
resolution of the dynamical stripping time. In contrast to the
stronger phase mixing in the leading arm, the dynamics of the
longer trailing arm demonstrate much better energy sorting of the
debris, so that position along the trailing tail is much better
correlated to the amount of time since a star was stripped from
the Sgr core. Further scrutiny of cleanly isolated trailing arm
stars may reveal even more clear chemical trends with dynamical
age. We will investigate this in future work.

\acknowledgements M.-Y.C. and S.R.M. acknowledge support from NSF
grants AST-0307851 and AST-0807945. This project was also
supported by the {\it SIM Lite} key project {\it Taking Measure of
the Milky Way} under NASA/JPL contract 1228235. VVS and KC also
thank support from the NSF via grant AST-0646790. D.G. gratefully
acknowledges support from the Chilean {\sl Centro de Astrof\'\i
sica} FONDAP No. 15010003 and the Chilean Centro de Excelencia en
Astrof\'\i sica y Tecnolog\'\i as Afines (CATA).

\begin{deluxetable}{lccc}
\tabletypesize{\scriptsize} \tablecaption{Atomic Data for The
Selected Lines } \tablewidth{0pt} \tablehead{ \colhead{} &
\colhead{$\lambda$} &\colhead{$\chi$} &\colhead{}
 \\
 \colhead{Ion} &\colhead{(\AA\ )} &\colhead{(eV)} &
 \colhead{\it gf}
 }

\startdata
\ion{Ti}{+1} & 7489.572 & 2.249 & 2.339E-01 \\
 & 7496.120 & 2.240 & 8.770E-02 \\
\ion{Y}{+2} & 7450.320 & 1.740 & 1.202E-01 \\
\ion{La}{+2} & 7483.234 & 0.126 & 2.080E-04 \\
 & 7483.237 & 0.126 & 4.160E-04 \\
 & 7483.259 & 0.126 & 2.080E-04 \\
 & 7483.262 & 0.126 & 4.160E-04 \\
 & 7483.267 & 0.126 & 6.240E-04 \\
 & 7483.304 & 0.126 & 4.160E-04 \\
 & 7483.308 & 0.126 & 6.240E-04 \\
 & 7483.315 & 0.126 & 8.320E-04 \\
 & 7483.367 & 0.126 & 6.240E-04 \\
 & 7483.374 & 0.126 & 8.320E-04 \\
 & 7483.382 & 0.126 & 1.039E-03 \\
 & 7483.449 & 0.126 & 8.320E-04 \\
 & 7483.458 & 0.126 & 1.039E-03 \\
 & 7483.468 & 0.126 & 1.247E-03 \\
 & 7483.550 & 0.126 & 1.039E-03 \\
 & 7483.561 & 0.126 & 1.247E-03 \\
 & 7483.573 & 0.126 & 1.455E-03 \\
 & 7483.670 & 0.126 & 1.247E-03 \\
 & 7483.682 & 0.126 & 1.455E-03 \\

\enddata
\end{deluxetable}

\begin{deluxetable}{lccc}
\tabletypesize{\scriptsize} \tablecaption{Equivalent Width
Measurements} \tablewidth{0pt} \tablehead{ \colhead{Star No.} &
\colhead{\ion{Ti}{+1}} &\colhead{\ion{Ti}{+1}}
&\colhead{\ion{Y}{+2}}
 \\
 \colhead{} &\colhead{7489.572} &\colhead{7496.120} &\colhead{7450.320}
 }

\startdata
Sgr(core) &  &  &  \\
$1849222-293217$ & \nodata & 193.5 & 38.3 \\
$1853333-320146$ & \nodata & \nodata & 14.1 \\
$1854283-295740$ & 147.1 & 124.8 & 35.1 \\
$1855341-302055$ & 134.0 & 95.8 & 38.9 \\
$1855556-293316$ & \nodata & \nodata & 28.5 \\
$1902135-313030$ & 64.9 & 49.7 & 34.8 \\
\\
Sgr (north leading arm)&  &  &  \\
$0919216+202305$ & 105.8 & 91.0 & 12.0 \\
$0925364+213807$ & 113.0 & 109.1 & 37.0 \\
$1034395+245206$ & 121.5 & 66.8 & 72.0 \\
$1100516+130216$ & 74.3 & 57.1 & 23.7 \\
$1101112+191311$ & 117.9 & 84.3 & 32.0 \\
$1111493+063915$ & 115.6 & 84.8 & 24.4 \\
$1112480+013211$ & 112.0 & 91.0 & 40.0 \\
$1114573-215126$ & 112.0 & 113.0 & 16.0 \\
$1116118-333057$ & \nodata & \nodata & 39.0 \\
$1128316-031647$ & 122.3 & 97.5 & 37.8 \\
$1135388-022602$ & 132.1 & 113.5 & 24.7 \\
$1140226-192500$ & 113.7 & 98.2 & 24.0 \\
$1208101-090753$ & 113.3 & 84.8 & 18.8 \\
$1223590-073028$ & 111.3 & 103.8 & 18.0 \\
$1224255-061852$ & 121.0 & 89.2 & 33.8 \\
$1227367-031834$ & 114.0 & 93.7 & 33.0 \\
$1236549-002941$ & 126.1 & 108.6 & 36.0 \\
$1249078+084455$ & 100.6 & 95.0 & 34.0 \\
$1318500+061112$ & 92.3 & 78.8 & 22.0 \\
$1319368-000817$ & 109.5 & 105.6 & 24.0 \\
$1330472-211847$ & 113.0 & 84.2 & 24.0 \\
$1334532+042053$ & 100.5 & 96.2 & 22.0 \\
$1348366+220101$ & 99.1 & 75.4 & 19.8 \\
$1407060+063311$ & 115.4 & 109.5 & 22.3 \\
$1411221-061013$ & 106.9 & 104.2 & 26.0 \\
$1435018+070827$ & 108.7 & 88.9 & 21.7 \\
$1450544+244357$ & 74.0 & 54.9 & 20.0 \\
$1456137+151112$ & 117.0 & 88.2 & 52.3 \\
$1512142-075250$ & 100.8 & 77.5 & 26.0 \\
$1538472+494218$ & 131.5 & 123.8 & 26.0 \\
\\
Sgr(south leading arm) &  &  &  \\
$2031334-324453$ & \nodata & 66.2 & 19.2 \\
$2037196-291738$ & 131.6 & 111.1 & 19.0 \\
$2046335-283547$ & \nodata & 63.3 & 12.8 \\
$2050020-345336$ & 99.6 & 79.5 & 24.3 \\
$2105585-275602$ & 129.9 & 82.0 & 24.5 \\
$2114412-301256$ & 126.4 & 100.2 & 40.8 \\
$2130445-210034$ & 87.7 & 51.6 & 21.4 \\
$2135183-203457$ & 167.0 & 134.6 & 35.5 \\
$2154471-224050$ & 127.4 & 123.6 & 31.9 \\
$2226328-340408$ & 53.7 & 38.8 & 20.5 \\
\\
NGP &  &  &  \\
$1033045+491604$ & 89.4 & 60.4 & 44.4 \\
$1041479+294917$ & 111.1 & 66.1 & 34.7 \\
$1051302+004400$ & 92.0 & 67.5 & 23.4 \\
$1115376+000800$ & 116.3 & 92.1 & 21.5 \\
$1214190+071358$ & 49.9 & 42.3 & 13.0 \\
$1257013+260046$ & 92.3 & 73.2 & 23.0 \\
$1343047+221636$ & 102.5 & 75.0 & 24.0 \\
$1412161+294303$ & 119.7 & 103.2 & 40.0 \\
$1424425+414932$ & 111.1 & 100.6 & 24.0 \\
$1429456+230043$ & 88.5 & 72.7 & 18.0 \\
$1513011+222640$ & 73.6 & 69.0 & 22.0 \\
$1536502+580017$ & 95.1 & 82.8 & 26.0 \\
$1545189+291310$ & 64.5 & 53.9 & 13.0 \\
\\
Standard Stars &  &  & \\
Arcturus & 74.2 & 52.9 & 27.9 \\
$\beta$ Peg & 131.0 & 106.0 & 34.0\\
$\beta$ And & 120.7 & 103.0 & 47.7 \\
$\rho$ Per & 127.7 & 111.4 & 40.4 \\
HD146051(mike) & 129.1 & 100.8 & 43.6 \\
HD146051(kpno) & 124.9 & 95.7 & 45.8 \\

\enddata

\end{deluxetable}
\clearpage

\begin{deluxetable}{lccccccccccccccc}
\tabletypesize{\scriptsize} \setlength{\tabcolsep}{0.02in}
\tablecaption{Chemical Abundances for The Program Stars}
\tablewidth{0pt} \tablehead{ \colhead{Star No.} &
 \colhead{$T_{\rm
eff}$} & \colhead{log {\it g}} & \colhead{$\xi$ }  &
\colhead{A(Fe)} & \colhead{[Fe/H]} & \colhead{standard} &
\colhead{A(Ti)} & \colhead{[Ti/Fe]} & \colhead{standard} &
\colhead{A(Y)} & \colhead{[Y/Fe]} &\colhead{standard} &
\colhead{A(La)} & \colhead{[La/Fe]} & \colhead{standard}
\\
\colhead{} & \colhead{Houdashelt (K)} & \colhead{(dex)} &
\colhead{(${\rm km\,s^{-1}}$)} & \colhead{} & \colhead{} &
\colhead{deviation} & \colhead{} & \colhead{} &
\colhead{deviation}& \colhead{} & \colhead{} & \colhead{deviation}
& \colhead{} & \colhead{} & \colhead{deviation}
 }

\startdata
Sun & \nodata & \nodata & \nodata & 7.45 & \nodata & \nodata & 4.90 & \nodata & \nodata & 2.21 & \nodata & \nodata & 1.13 & \nodata & \nodata \\
\\
Sgr(core) &  &  &  &  &  &  &  &  &  &  &  &  &  &  &  \\
$1849222-293217$ & 3850 & 0.9 & 2.43 & 7.24 & -0.21 & 0.10 & 5.54 & 0.85 & \nodata & 1.70 & -0.30 & \nodata & 1.51 & 0.59 & 0.02 \\
$1853333-320146$ & 3750 & 0.7 & 2.60 & 7.15 & -0.30 & 0.14 & CR\tablenotemark{a} & \nodata & \nodata & 1.10 & -0.81 & \nodata & 1.38 & 0.55 & 0.01 \\
$1854283-295740$ & 3750 & 0.0(-) & 3.21 & 6.48 & -0.97 & 0.06 & 4.20 & 0.27 & 0.17 & 0.98 & -0.26 & \nodata & 0.65 & 0.49 & 0.02 \\
$1855341-302055$ & 3800 & 1.0 & 1.84 & 7.47 & 0.02 & 0.08 & 4.66 & -0.32 & 0.09 & 1.89 & -0.34 & \nodata & 1.36 & 0.21 & 0.02 \\
$1855556-293316$ & 3700 & 0.5 & 2.36 & 7.18 & -0.27 & 0.07 & CR\tablenotemark{a} & \nodata & \nodata & 1.37 & -0.57 & \nodata & 1.37 & 0.51 & 0.03 \\
$1902135-313030$ & 3750 & 0.0(-) & 1.04 & 6.41 & -1.04 & 0.11 & 3.80 & -0.06 & 0.10 & 1.18 & 0.01 & \nodata & CR\tablenotemark{a} & \nodata & \nodata \\
\\
Sgr (north leading arm)&  &  &  &  &  &  &  &  &  &  &  &  &  &  &  \\
$0919216+202305$ & 3700 & 0.25 & 1.47 & 6.82 & -0.63 & 0.08 & 4.29 & 0.02 & 0.12 & 0.75 & -0.83 & 0.06 & 0.14 & -0.36 & 0.00 \\
$0925364+213807$ & 3600 & 0.5 & 1.29 & 7.22 & -0.23 & 0.07 & 4.70 & 0.03 & 0.24 & 1.76 & -0.22 & 0.07 & CR\tablenotemark{a} & \nodata & \nodata \\
$1034395+245206$ & 3700 & 0.25 & 1.45 & 6.80 & -0.65 & 0.09 & 4.25 & 0.00 & 0.33 & 2.08 & 0.52 & 0.12 & CR\tablenotemark{a} & \nodata & \nodata \\
$1100516+130216$ & 3800 & 0.0 & 1.35 & 6.39 & -1.06 & 0.08 & 3.93 & 0.09 & 0.11 & 0.82 & -0.33 & 0.13 & 0.06 & -0.01 & 0.03 \\
$1101112+191311$ & 3700 & 0.8 & 1.51 & 7.47 & 0.02 & 0.09 & 4.53 & -0.39 & 0.11 & 1.78 & -0.45 & 0.09 & 1.15 & 0.00 & 0.04 \\
$1111493+063915$ & 3600 & 0.0(-) & 1.71 & 6.75 & -0.70 & 0.09 & 4.08 & -0.12 & 0.02 & 1.03 & -0.48 & \nodata & 0.05 & -0.38 & 0.04 \\
$1112480+013211$ & 3800 & 0.5 & 1.60 & 6.97 & -0.48 & 0.12 & 4.42 & 0.00 & 0.07 & 1.55 & -0.18 & 0.02 & 0.44 & -0.21 & 0.02 \\
$1114573-215126$ & 3550 & 0.0(-) & 1.33 & 6.64 & -0.81 & 0.07 & 4.42 & 0.33 & 0.30 & 0.81 & -0.59 & 0.05 & CR\tablenotemark{a} & \nodata & \nodata \\
$1116118-333057$ & 3650 & 0.0(-) & 1.39 & 6.32 & -1.13 & 0.05 & CR\tablenotemark{a} & \nodata & \nodata & 0.19 & 0.19 & \nodata & CR\tablenotemark{a} & \nodata & \nodata \\
$1128316-031647$ & 3700 & 0.9 & 1.64 & 7.41 & -0.04 & 0.05 & 4.61 & -0.25 & 0.01 & 1.90 & -0.27 & \nodata & 1.10 & 0.01 & 0.06 \\
$1135388-022602$ & 3700 & 0.9 & 1.23 & 7.45 & 0.00 & 0.07 & 5.15 & 0.25 & 0.02 & 1.66 & -0.55 & \nodata & 0.75 & -0.38\tablenotemark{d} & 0.12 \\
$1140226-192500$ & 3800 & 0.6 & 1.16 & 7.07 & -0.38 & 0.05 & 4.78 & 0.26 & 0.08 & 1.31 & -0.52 & 0.08 & 0.53 & -0.22 & 0.05 \\
$1208101-090753$ & 3750 & 0.0(-) & 1.82 & 6.46 & -0.99 & 0.07 & 4.16 & 0.25 & 0.04 & 0.69 & -0.53 & \nodata & -0.16 & -0.30 & 0.05 \\
$1223590-073028$ & 3600 & 0.0 & 1.50 & 6.73 & -0.72 & 0.08 & 4.29 & 0.11 & 0.20 & 0.88 & -0.61 & 0.02 & CR\tablenotemark{a} & \nodata & \nodata \\
$1224255-061852$ & 3750 & 0.3 & 1.71 & 6.80 & -0.65 & 0.08 & 4.33 & 0.08 & 0.03 & 1.30 & -0.26 & \nodata & 0.19 & -0.29 & 0.06 \\
$1227367-031834$ & 3850 & 0.5 & 1.68 & 6.90 & -0.55 & 0.08 & 4.47 & 0.12 & 0.09 & 1.36 & -0.30 & 0.02 & 0.37 & -0.21 & 0.02 \\
$1236549-002941$ & 3750 & 0.5 & 1.33 & 7.06 & -0.39 & 0.10 & 4.80 & 0.29 & 0.05 & 1.53 & -0.28 & 0.03 & CR\tablenotemark{a} & \nodata & \nodata \\
$1249078+084455$ & 3800 & 0.3 & 1.52 & 6.78 & -0.67 & 0.10 & 4.37 & 0.14 & 0.22 & 1.31 & -0.23 & 0.13 & 0.38\tablenotemark{b} & -0.08 & 0.08 \\
$1318500+061112$ & 3850 & 0.4 & 1.67 & 6.68 & -0.77 & 0.10 & 4.21 & 0.08 & 0.16 & 1.01 & -0.43 & 0.14 & CR\tablenotemark{a} & \nodata & \nodata \\
$1319368-000817$ & 3500 & 0.0(-) & 1.64 & 6.86 & -0.59 & 0.08 & 4.16 & -0.15 & 0.25 & 1.09 & -0.53 & 0.02 & CR\tablenotemark{a} & \nodata & \nodata \\
$1330472-211847$ & 3850 & 0.0(-) & 1.75 & 6.35 & -1.10 & 0.08 & 4.34 & 0.54 & 0.03 & 0.76 & -0.35 & 0.09 & CR\tablenotemark{a} & \nodata & \nodata \\
$1334532+042053$ & 3700 & 0.25 & 1.49 & 6.83 & -0.62 & 0.06 & 4.28 & 0.00 & 0.24 & 1.05 & -0.54 & 0.15 & 0.29 & -0.22 & 0.03 \\
$1348366+220101$ & 3800 & 0.1 & 1.49 & 6.63 & -0.82 & 0.09 & 4.20 & 0.12 & 0.05 & 0.80 & -0.59 & \nodata & -0.01 & -0.32\tablenotemark{d} & 0.12 \\
$1407060+063311$ & 3700 & 0.0(-) & 1.78 & 6.50 & -0.95 & 0.09 & 4.29 & 0.34 & 0.23 & 0.84 & -0.42 & \nodata & -0.06 & -0.24 & 0.05 \\
$1411221-061013$ & 3700 & 0.25 & 1.51 & 6.89 & -0.56 & 0.06 & 4.38 & 0.04 & 0.26 & 1.18 & -0.47 & 0.04 & 0.10 & -0.47 & 0.05 \\
$1435018+070827$ & 3700 & 0.0(-) & 1.61 & 6.53 & -0.92 & 0.09 & 4.18 & 0.20 & 0.09 & 0.83 & -0.46 & \nodata & -0.19 & -0.40\tablenotemark{d} & 0.14 \\
$1450544+244357$ & 3800 & 0.0 & 1.63 & 6.37 & -1.08 & 0.08 & 3.86 & 0.04 & 0.11 & 0.70 & -0.43 & 0.19 & -0.10 & -0.15 & 0.01 \\
$1456137+151112$ & 3750 & 0.0(-) & 1.71 & 6.47 & -0.98 & 0.08 & 4.26 & 0.34 & 0.02 & 1.38 & 0.15 & \nodata & -0.07 & -0.19\tablenotemark{d} & 0.10 \\
$1512142-075250$ & 3700 & 0.0(-) & 1.26 & 6.48 & -0.97 & 0.08 & 4.19 & 0.26 & 0.01 & 0.97 & -0.27 & \nodata & CR\tablenotemark{a} & \nodata & \nodata \\
$1538472+494218$ & 3600 & 0.0(-) & 1.52 & 6.39 & -1.06 & 0.08 & 4.55 & 0.71 & 0.20 & 0.99 & -0.16 & 0.01 & -0.20 & -0.27 & 0.06 \\
\\
Sgr(south leading arm) &  &  &  &  &  &  &  &  &  &  &  &  &  &  &  \\
$2031334-324453$ & 3800 & 0.0(-) & 2.67 & 6.13 & -1.32 & 0.11 & 3.97 & 0.39 & \nodata & 0.54 & -0.35 & \nodata & CR\tablenotemark{a} & \nodata & \nodata \\
$2037196-291738$ & 3700 & 0.0 & 2.32 & 6.75 & -0.70 & 0.09 & 4.21 & 0.01 & 0.14 & 0.79 & -0.72 & \nodata & 0.39 & -0.04 & 0.02 \\
$2046335-283547$ & 3750 & 0.0(-) & 2.56 & 6.19 & -1.26 & 0.06 & 3.86 & 0.22 & \nodata & 0.42 & -0.53 & \nodata & CR\tablenotemark{a} & \nodata & \nodata \\
$2050020-345336$ & 3800 & 0.0(-) & 2.12 & 6.41 & -1.04 & 0.10 & 4.06 & 0.20 & 0.14 & 0.79 & -0.38 & \nodata & -0.04 & -0.13 & 0.02 \\
$2105585-275602$ & 3700 & 0.0(-) & 2.13 & 6.49 & -0.96 & 0.10 & 4.09 & 0.15 & 0.08 & 0.87 & -0.38 & \nodata & CR\tablenotemark{a} & \nodata & \nodata \\
$2114412-301256$ & 3750 & 0.0(-) & 2.06 & 6.30 & -1.15 & 0.10 & 4.25 & 0.50 & 0.08 & 1.12 & 0.06 & \nodata & 0.21 & 0.23 & 0.01 \\
$2130445-210034$ & 3750 & 0.0(-) & 2.62 & 6.10 & -1.35 & 0.10 & 3.71 & 0.16 & 0.04 & 0.63 & -0.23 & \nodata & 0.04 & 0.26 & 0.03 \\
$2135183-203457$ & 3700 & 0.0(-) & 2.30 & 6.55 & -0.90 & 0.13 & 4.52 & 0.52 & 0.3 & 1.08 & -0.23 & \nodata & 0.04 & -0.19 & 0.01 \\
$2154471-224050$ & 3800 & 0.1 & 2.14 & 6.54 & -0.91 & 0.06 & 4.44 & 0.45 & 0.26 & 1.00 & -0.30 & \nodata & 0.19 & -0.03 & 0.04 \\
$2226328-340408$ & 3800 & 0.0(-) & 1.93 & 6.11 & -1.34 & 0.09 & 3.58 & 0.02 & 0.15 & 0.61 & -0.26 & \nodata & -0.47 & -0.26 & 0.01 \\
\\
NGP &  &  &  &  &  &  &  &  &  &  &  &  &  &  &  \\
$1033045+491604$ & 3800 & 0.3 & 1.56 & 6.70 & -0.75 & 0.09 & 4.01 & -0.14 & 0.00 & 1.47 & 0.01 & 0.14 & 0.46 & 0.08 & 0.01 \\
$1041479+294917$ & 3800 & 0.0(-) & 1.56 & 6.21 & -1.24 & 0.10 & 4.20 & 0.54 & 0.15 & 1.01 & 0.04 & \nodata & 0.04 & 0.15 & 0.01 \\
$1051302+004400$ & 3800 & 0.0(-) & 1.65 & 6.07 & -1.38 & 0.06 & 4.05 & 0.53 & 0.06 & 0.70 & -0.13 & \nodata & -0.32 & -0.07 & \nodata \\
$1115376+000800$ & 3800 & 0.0 & 1.56 & 6.49 & -0.96 & 0.10 & 4.42 & 0.48 & 0.04 & 0.76 & -0.49 & \nodata & -0.06 & -0.23\tablenotemark{d} & 0.09 \\
$1214190+071358$ & 3750 & 0.0(-) & 1.73 & 6.32 & -1.13 & 0.10 & 3.51 & -0.26 & 0.21 & 0.45 & -0.63 & 0.18 & CR\tablenotemark{a} & \nodata & \nodata \\
$1257013+260046$ & 3750 & 0.0(-) & 1.68 & 6.49 & -0.96 & 0.09 & 4.01 & 0.07 & 0.11 & 0.81 & -0.44 & 0.22 & 0.01 & -0.16 & 0.04 \\
$1343047+221636$ & 3750 & 0.0(-) & 2.26 & 6.37 & -1.08 & 0.10 & 3.94 & 0.12 & 0.09 & 0.81 & -0.32 & 0.02 & -0.03 & -0.08 & 0.03 \\
$1412161+294303$ & 3800 & 0.6 & 1.76 & 7.02 & -0.43 & 0.07 & 4.50 & 0.03 & 0.12 & 1.59 & -0.19 & 0.05 & 0.49 & -0.21 & 0.06 \\
$1424425+414932$ & 3700 & 0.1 & 1.59 & 6.73 & -0.72 & 0.07 & 4.31 & 0.13 & 0.17 & 1.01 & -0.48 & 0.06 & 0.10 & -0.31 & 0.02 \\
$1429456+230043$ & 3750 & 0.0(-) & 1.88 & 6.48 & -0.97 & 0.11 & 3.93 & 0.00 & 0.16 & 0.66 & -0.58 & 0.09 & 0.04 & -0.12 & 0.01 \\
$1513011+222640$ & 3700 & 0.0(-) & 1.17 & 6.60 & -0.85 & 0.10 & 3.94 & -0.11 & 0.24 & 0.91 & -0.45 & 0.08 & -0.12\tablenotemark{c} & -0.40 & 0.02 \\
$1536502+580017$ & 3750 & 0.0(-) & 1.53 & 6.61 & -0.84 & 0.10 & 4.14 & 0.08 & 0.17 & 0.94 & -0.43 & 0.08 & -0.07 & -0.36 & 0.01 \\
$1545189+291310$ & 3850 & 0.1 & 1.52 & 6.45 & -1.00 & 0.09 & 3.96 & 0.06 & 0.07 & 0.51 & -0.70 & 0.04 & -0.18 & -0.31 & 0.03 \\
\\
Standard Stars &  &  &  &  &  &  &  &  &  &  &  &  &  &  &  \\
Arcturus & 4250 & 1.4 & 1.66 & 6.74 & -0.71 & 0.10 & 4.54 & 0.35 & 0.07 & 1.44 & -0.06 & 0.09 & 0.34 & -0.08 & 0.07 \\
$\beta$ Peg & 3750 & 0.6 & 1.73 & 6.98 & -0.47 & 0.08 & 4.59 & 0.16 & 0.02 & 1.54 & -0.20 & 0.02 & 0.57 & -0.09 & 0.01 \\
$\beta$ And & 3850 & 0.9 & 1.96 & 7.12 & -0.33 & 0.06 & 4.55 & -0.02 & 0.12 & 1.89 & 0.01 & 0.00 & 0.79 & -0.01 & 0.02 \\
$\rho$ Per & 3650 & 0.7 & 1.35 & 7.36 & -0.09 & 0.15 & 4.93 & 0.12 & 0.06 & 1.97 & -0.15 & 0.10 & 0.86 & -0.18 & 0.03 \\
HD146051(mike) & 4000 & 1.0 & 2.07 & 6.98 & -0.47 & 0.08 & 4.70 & 0.27 & 0.05 & 1.69 & -0.05 & \nodata & 0.75 & 0.09 & 0.02 \\
HD146051(kpno) & 4000 & 1.0 & 1.81 & 7.06 & -0.39 & 0.08 & 4.74 & 0.23 & 0.01 & 1.76 & -0.06 & 0.06 & 0.80 & 0.06 & 0.02 \\
HD146051 (average)& 4000 & 1.0 & \nodata & 7.02 & -0.43 & \nodata & 4.72 & 0.25 & \nodata & 1.73 & -0.06 & \nodata & 0.78 & 0.08 & \nodata \\

\enddata

\tablenotetext{a}{~Lines unmeasurable due to cosmic rays or other
defects are marked ''CR''.} \tablenotetext{b}{~Measurement
uncertain due to spectrum defect on the blue edge of the observed
La line.} \tablenotetext{c}{~Measurement uncertain due to unusual
shape of the observed La line.} \tablenotetext{d}{~Larger
measurement uncertainty of La due to noisier spectrum. These
measurements were excluded from the figures. }

\end{deluxetable}
\clearpage

\begin{deluxetable}{lccccccc}
\tabletypesize{\scriptsize} \tablecaption{Comparison of The
Control Sample Stars with References} \tablewidth{0pt} \tablehead{
\colhead{Star Name} &
 \colhead{$T_{\rm
eff}$} & \colhead{log {\it g}} & \colhead{$\xi$ }  &
\colhead{A(Fe)} & \colhead{A(Ti)} & \colhead{A(Y)} &
\colhead{A(La)}
\\
\colhead{Reference} & \colhead{(K)} & \colhead{(dex)} &
\colhead{(${\rm km\,s^{-1}}$)} & \colhead{} & \colhead{} &
\colhead{} & \colhead{}
 }

\startdata
Arcturus &  &  &  &  &  &  & \\
This Work & 4250 & 1.4 & 1.66 & 6.74$\pm$0.10 & 4.54$\pm$0.07 & 1.44$\pm$0.09 & 0.34$\pm$0.07 \\
McWilliam \& Rich 1994 & 4280 & 1.3 & 1.4 & 6.98$\pm$0.18 & 4.77$\pm$0.13 & 1.28$\pm$0.30 & 0.53$\pm$0.01 \\
Smith et al.\ 2000 & 4300 & 1.7 & 1.6 & 6.78$\pm$0.11 & 4.64$\pm$0.12 & 1.4$\pm$0.15 & 0.62 \\
\\
$\beta$ Peg &  &  &  &  &  &  & \\
This Work & 3750 & 0.6 & 1.73 & 6.98$\pm$0.08 & 4.59$\pm$0.02 & 1.54$\pm$0.02 & 0.57$\pm$0.01 \\
Smith \& Lambert 1985\tablenotemark{a} & 3600 & 1.2 & 2.0 & 7.41$\pm$0.13 & 4.87$\pm$0.10 & 2.19 & \nodata \\
This Work with S\&L Parameters & 3600 & 1.2 & 2.0 & 7.32$\pm$0.10 & 4.59$\pm$0.04 & 2.00$\pm$0.02 & \nodata \\
\\
$\beta$ And &  &  &  &  &  &  & \\
This Work & 3850 & 0.9 & 1.96 & 7.12$\pm$0.06 & 4.55$\pm$0.12 & 1.89$\pm$0.01 & 0.79$\pm$0.02 \\
Smith \& Lambert 1985\tablenotemark{a} & 3800 & 1.6 & 2.1 & 7.42$\pm$0.11 & 4.87$\pm$0.12 & 2.35 & \nodata \\
This Work with S\&L Parameters & 3800 & 1.6 & 2.1 & 7.38$\pm$0.07 & 4.62$\pm$0.13 & 2.29$\pm$0.00 & \nodata \\
\\
$\rho$ Per &  &  &  &  &  &  & \\
This Work & 3650 & 0.7 & 1.35 & 7.36$\pm$0.15 & 4.93$\pm$0.06 & 1.97$\pm$0.10 & 0.86$\pm$0.03 \\
Smith \& Lambert 1986\tablenotemark{a} & 3500 & 0.8 & 1.8 & 7.57$\pm$0.17 & 4.86$\pm$0.14 & 2.00 & \nodata \\
This Work with S\&L Parameters & 3500 & 0.8 & 1.8 & 7.30$\pm$0.20 & 4.58$\pm$0.12 & 2.00$\pm$0.09 & \nodata \\

\enddata

\tablenotetext{a}{~$A$(X) are derived using the relative abundance
ratios [X/H] in S\&L and the absolute abundances of $\alpha$ Tau.
The absolute abundances of $\alpha$ Tau are derived from the
stellar parameters and the EWs in S\&L, but with the Kurucz model
atmospheres and the $gf$-values in this work. }

\end{deluxetable}

\begin{deluxetable}{lccc}
\tabletypesize{\scriptsize} \tablecaption{Sensitivity of
Abundances to Stellar Parameters} \tablewidth{0pt} \tablehead{
\colhead{Star Name} &
 \colhead{$\Delta T_{\rm
eff}=+100$} & \colhead{$\Delta$log {\it g}$=+0.2$} &
\colhead{$\Delta\xi=+0.2$ }
\\
\colhead{Element} &
 \colhead{(K)} & \colhead{(dex)} & \colhead{(${\rm km\,s^{-1}}$) } }

\startdata

$\beta$ Peg & & & \\
$\Delta A$(Fe) & $-0.06$ & $+0.08$ & $-0.09$ \\
$\Delta A$(Ti) & $+0.10$ & $+0.03$ & $-0.10$ \\
$\Delta A$(Y) & $-0.03$ & $+0.09$ & $-0.02$ \\
$\Delta A$(La) & $+0.03$ & $+0.08$ & $+0.00$ \\
\\
$\beta$ And & & & \\
$\Delta A$(Fe) & $-0.07$ & $+0.07$ & $-0.09$ \\
$\Delta A$(Ti) & $+0.10$ & $+0.03$ & $-0.08$ \\
$\Delta A$(Y) & $-0.04$ & $+0.09$ & $-0.02$ \\
$\Delta A$(La) & $+0.01$ & $+0.08$ & $-0.01$ \\
\\
$\rho$ Per & & & \\
$\Delta A$(Fe) & $-0.12$ & $+0.06$ & $-0.09$ \\
$\Delta A$(Ti) & $+0.04$ & $+0.04$ & $-0.15$ \\
$\Delta A$(Y) & $-0.07$ & $+0.08$ & $-0.04$ \\
$\Delta A$(La) & $+0.00$ & $+0.09$ & $+0.00$ \\

\enddata

\end{deluxetable}

\clearpage
\begin{deluxetable}{lccccc}
\tabletypesize{\scriptsize} \tablecaption{Score Card for The
Program Stars} \tablewidth{0pt} \tablehead{ \colhead{Star No.} &
\colhead{N$_\sigma({\rm [Ti/Fe]})$} & \colhead{N$_\sigma({\rm
[Y/Fe]})$} & \colhead{N$_\sigma({\rm [La/Fe]})$} & \colhead{Sum of
Abs(N$_\sigma$)} & \colhead{Average of Abs(N$_\sigma$)}
\\
\colhead{} & \colhead{} & \colhead{} & \colhead{} & \colhead{} &
\colhead{}
 }

\startdata
Sgr(core) &  &  &  &  &  \\
$1849222-293217$ & 15.6 & -2.3 & 7.8 & 25.7 & 8.6 \\
$1853333-320146$ & \nodata & -5.7 & 6.8 & 12.5 & 6.3 \\
$1854283-295740$ & 0.1 & -1.4 & 5.1 & 6.6 & 2.2 \\
$1855341-302055$ & -5.5 & -2.7 & 3.7 & 11.9 & 4.0 \\
$1855556-293316$ & \nodata & -4.1 & 6.0 & 10.1 & 5.1 \\
$1902135-313030$ & -3.1 & 0.5 & \nodata & 3.6 & 1.8 \\
\\
Sgr (north leading arm)&  &  &  &  &  \\
$0919216+202305$ & -0.9 & -5.5 & -1.9 & 8.3 & 2.8 \\
$0925364+213807$ & -0.3 & -1.7 & \nodata & 2.0 & 1.0 \\
$1034395+245206$ & -1.3 & 3.7 & \nodata & 5.0 & 2.5 \\
$1100516+130216$ & -1.7 & -1.8 & 0.5 & 4.0 & 1.3 \\
$1101112+191311$ & -6.8 & -3.5 & 1.8 & 12.1 & 4.0 \\
$1111493+063915$ & -3.7 & -3.1 & -2.2 & 9.0 & 3.0 \\
$1112480+013211$ & -0.9 & -1.2 & -0.5 & 2.6 & 0.9 \\
$1114573-215126$ & 0.6 & -3.8 & \nodata & 4.4 & 2.2 \\
$1116118-333057$ & \nodata & 1.8 & \nodata & 1.8 & 1.8 \\
$1128316-031647$ & -4.4 & -2.2 & 2.0 & 8.6 & 2.9 \\
$1135388-022602$ & 5.0 & -4.1 & -1.3 & 10.4 & 3.5 \\
$1140226-192500$ & 4.2 & -3.6 & -0.5 & 8.3 & 2.8 \\
$1208101-090753$ & -0.1 & -3.2 & -1.8 & 5.1 & 1.7 \\
$1223590-073028$ & -1.5 & -4.0 & \nodata & 5.5 & 2.8 \\
$1224255-061852$ & 0.2 & -1.6 & -1.3 & 3.1 & 1.0 \\
$1227367-031834$ & 1.2 & -2.0 & -0.4 & 3.6 & 1.2 \\
$1236549-002941$ & 4.8 & -2.0 & \nodata & 6.8 & 3.4 \\
$1249078+084455$ & 1.3 & -1.4 & 0.5 & 3.2 & 1.1 \\
$1318500+061112$ & -1.8 & -2.7 & \nodata & 4.5 & 2.3 \\
$1319368-000817$ & -3.9 & -3.5 & \nodata & 7.4 & 3.7 \\
$1330472-211847$ & 2.7 & -1.9 & \nodata & 4.6 & 2.3 \\
$1334532+042053$ & -1.2 & -3.6 & -0.7 & 5.5 & 1.8 \\
$1348366+220101$ & -1.4 & -3.8 & -1.8 & 7.0 & 2.3 \\
$1407060+063311$ & 0.7 & -2.5 & -1.3 & 4.5 & 1.5 \\
$1411221-061013$ & -0.3 & -3.1 & -2.8 & 6.2 & 2.1 \\
$1435018+070827$ & -0.6 & -2.8 & -3.0 & 6.4 & 2.1 \\
$1450544+244357$ & -2.2 & -2.5 & -0.6 & 5.3 & 1.8 \\
$1456137+151112$ & 0.7 & 1.4 & -1.1 & 3.2 & 1.1 \\
$1512142-075250$ & 0.0 & -1.5 & \nodata & 1.5 & 0.8 \\
$1538472+494218$ & 4.3 & -0.6 & -1.6 & 6.5 & 2.2 \\
\\
Sgr(south leading arm) &  &  &  &  &  \\
$2031334-324453$ & 1.2 & -1.7 & \nodata & 2.9 & 1.5 \\
$2037196-291738$ & -1.2 & -4.7 & 0.8 & 6.7 & 2.2 \\
$2046335-283547$ & -0.4 & -3.0 & \nodata & 3.4 & 1.7 \\
$2050020-345336$ & -0.6 & -2.2 & -0.3 & 3.1 & 1.0 \\
$2105585-275602$ & -1.1 & -2.2 & \nodata & 3.3 & 1.7 \\
$2114412-301256$ & 2.3 & 0.9 & 2.7 & 5.9 & 2.0 \\
$2130445-210034$ & -1.0 & -0.9 & 2.7 & 4.6 & 1.5 \\
$2135183-203457$ & 2.5 & -1.2 & -0.7 & 4.4 & 1.5 \\
$2154471-224050$ & 1.8 & -1.7 & 0.7 & 4.2 & 1.4 \\
$2226328-340408$ & -2.4 & -1.1 & -1.8 & 5.3 & 1.8 \\
\\
NGP &  &  &  &  &  \\
$1033045+491604$ & -3.9 & 0.3 & 1.8 & 6.1 & 2.0 \\
$1041479+294917$ & 2.7 & 0.9 & 1.6 & 5.2 & 1.7 \\
$1051302+004400$ & 2.6 & -0.2 & -1.5 & 4.3 & 1.4 \\
$1115376+000800$ & 2.1 & -3.0 & -1.1 & 6.2 & 2.1 \\
$1214190+071358$ & -5.1 & -3.8 & \nodata & 8.9 & 4.5 \\
$1257013+260046$ & -1.9 & -2.6 & -0.5 & 5.0 & 1.7 \\
$1343047+221636$ & -1.4 & -1.7 & 0.0 & 3.1 & 1.0 \\
$1412161+294303$ & -0.2 & -1.3 & -0.4 & 1.9 & 0.6 \\
$1424425+414932$ & -1.3 & -3.1 & -1.6 & 6.0 & 2.0 \\
$1429456+230043$ & -2.6 & -3.6 & -0.2 & 6.4 & 2.1 \\
$1513011+222640$ & -3.6 & -2.8 & -2.5 & 8.9 & 3.0 \\
$1536502+580017$ & -1.8 & -2.7 & -1.9 & 6.4 & 2.1 \\
$1545189+291310$ & -2.0 & -4.4 & -1.9 & 8.3 & 2.8 \\

\enddata

\end{deluxetable}

\begin{figure}
\plotone{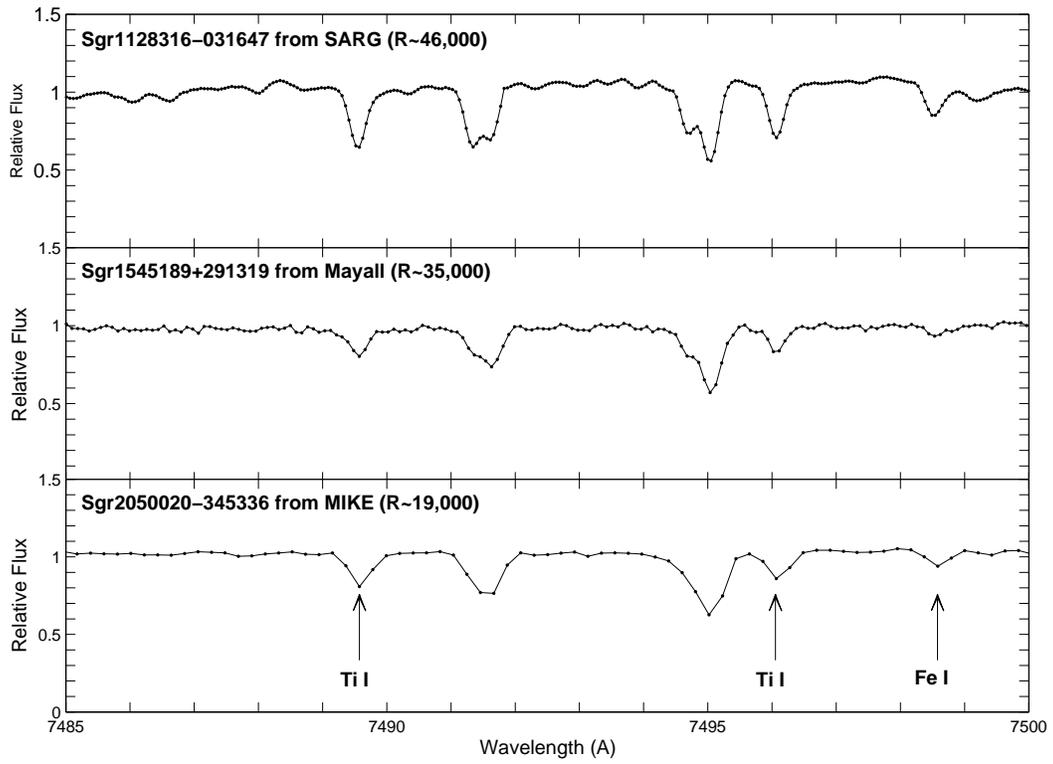} \caption{ Sample spectra of three M giants from
the three different spectrographs used in this study. Sample
titanium and iron lines are identified in the figure.
 }
\end{figure}

\begin{figure}
\includegraphics[angle=-90,scale=0.6]{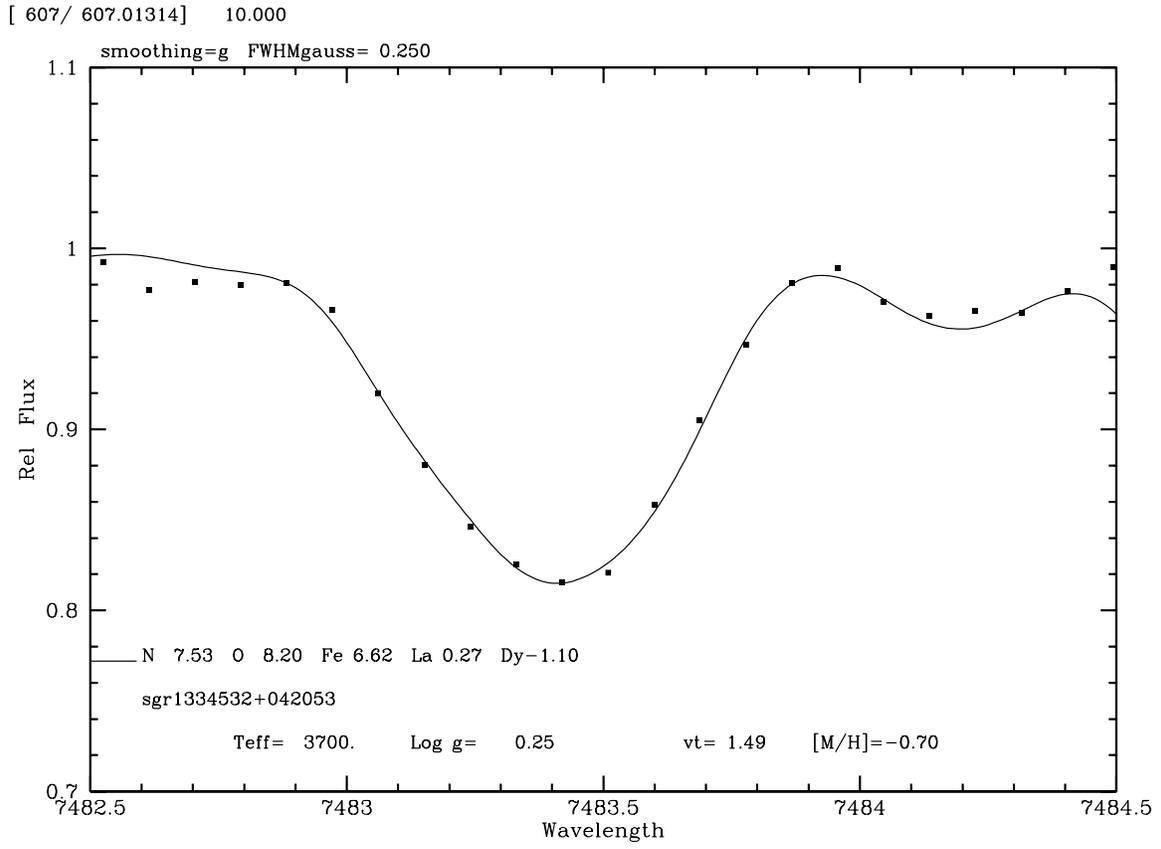}  \caption{
Sample observed (filled circles) and synthetic spectra (continuous
curves) for the \ion{La}{+2} 7483.5 \AA\ line in
Sgr1334532+042053.
 }
\end{figure}

\begin{figure}
\plotone{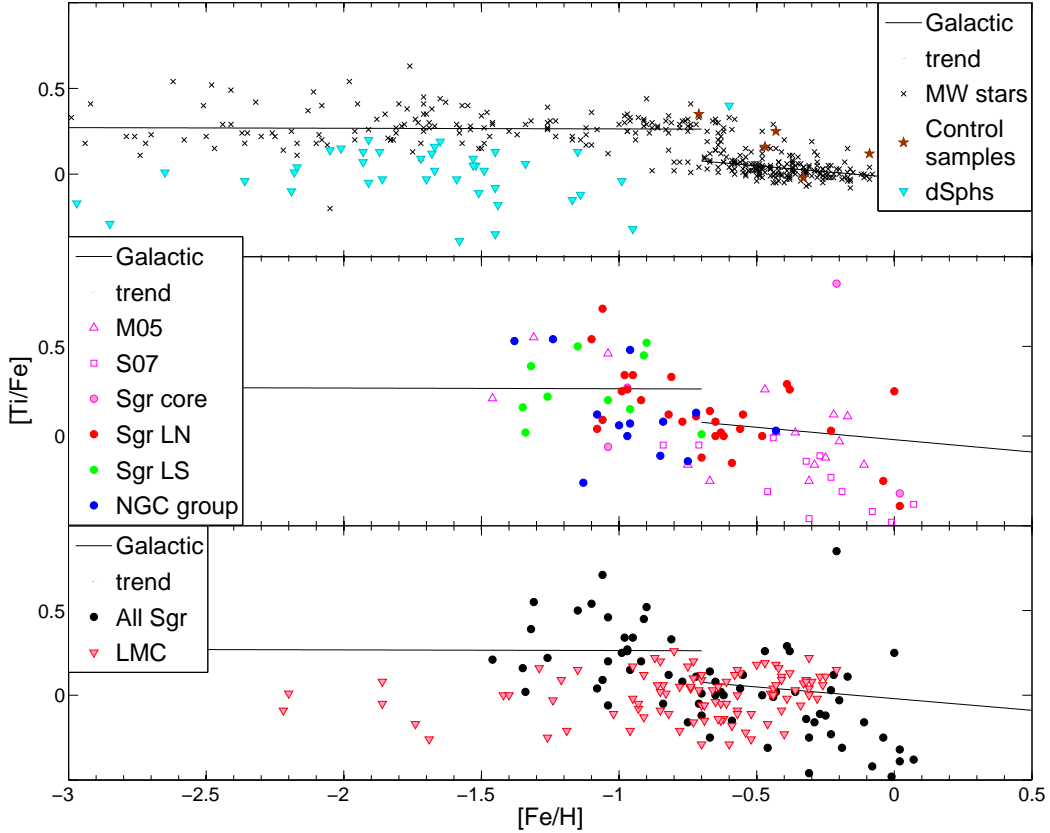} \caption{The distribution of [Ti/Fe] as a
function of [Fe/H] for ({\it top panel}) Milky Way (black crosses)
and dSph stars (cyan triangles). The MW data are from Fulbright
(2000), Johnson (2002) and Reddy et al.\ (2003). The dSph data are
from Shetrone et al.\ (2001; 2003), Sadakane et al.\ (2004) and
Geisler et al.\ (2005). The brown star symbols show the ``control
sample'' stars for our survey. The lines represent a two-piece
linear fit to the MW star distribution, where separate fits are
used to either side of the apparent transition at [Fe/H]=$-0.7$.
({\it middle panel}) Sgr stars with all stars in our spectroscopic
sample shown as filled, colored circles: Sgr core (magenta),
leading arm north (Sgr LN, red), leading arm south (Sgr LS, green)
and the ``NGC'' group of stars having more positive GSR radial
velocities that are well off the main leading arm velocity trend
(blue). The open symbols show Sgr core stars from Monaco et al.\
(2005b, M05) and Sbordone et al.\ (2007, S07). ({\it bottom
panel}) The pink triangles are LMC stars from Johnson et al.\
(2006), Pomp{\'e}ia et al.\ (2008) and Mucciarelli et al.\ (2008),
to which we compare the Sgr stars from the middle panel, shown
here as solid black circles. }
\end{figure}

\begin{figure}
\plotone{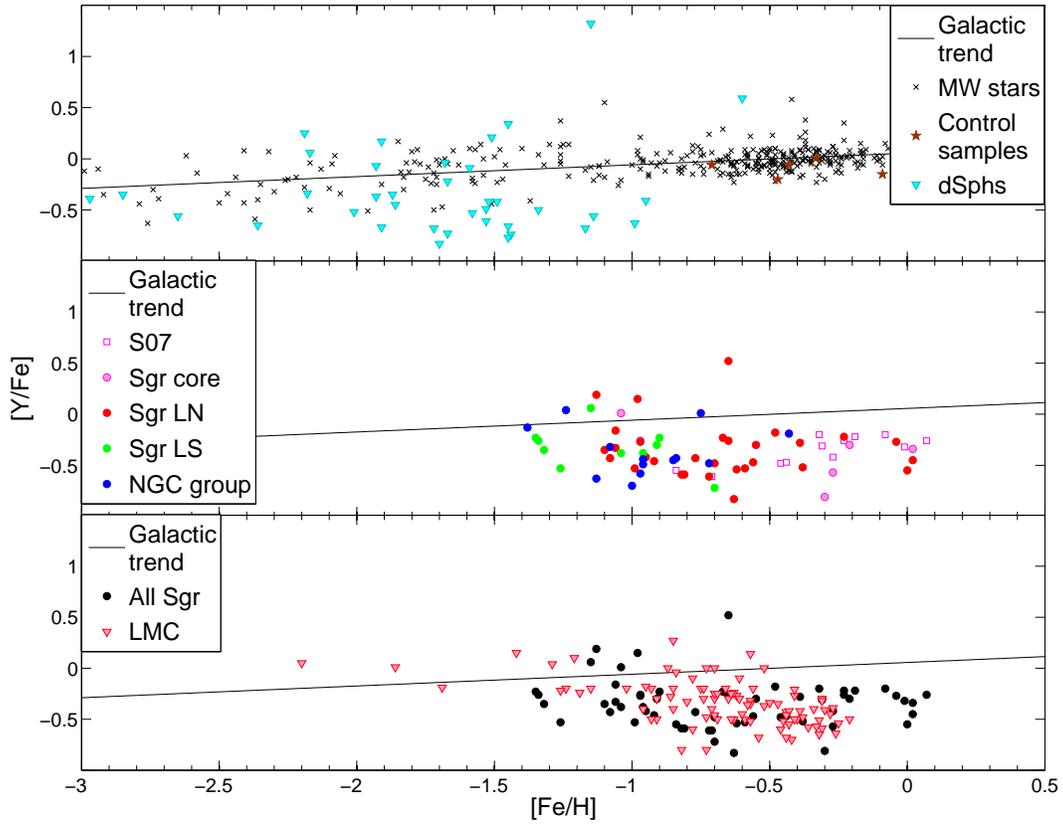} \caption{The same as Fig.\ 3, but for [Y/Fe]
versus [Fe/H]. The MW data are those published by Gratton \&
Sneden (1994), Fulbright (2000), Johnson (2002) and Reddy et al.\
(2003). Other data come from the same references as Fig.\ 3.}
\end{figure}

\begin{figure}
\plotone{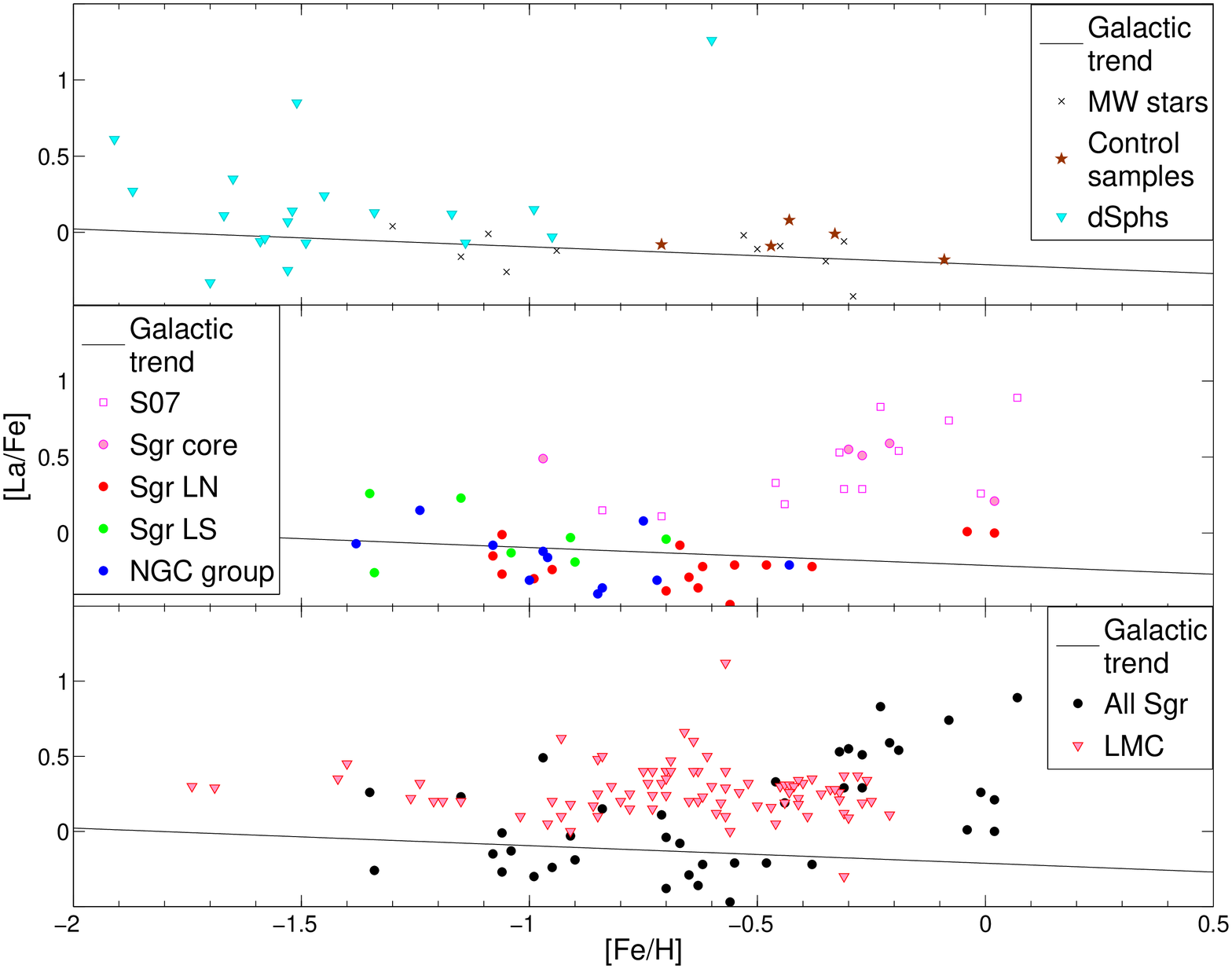} \caption{The same as Fig.\ 4, but for [La/Fe]
versus [Fe/H]. The data for Galactic stars come from Gratton \&
Sneden (1994), and for dSphs come from Shetrone et al.\ (2003),
Sadakane et al.\ (2004) and Geisler et al.\ (2005). The LMC data
come from the same references as Fig.\ 3.}
\end{figure}

\begin{figure}
\plotone{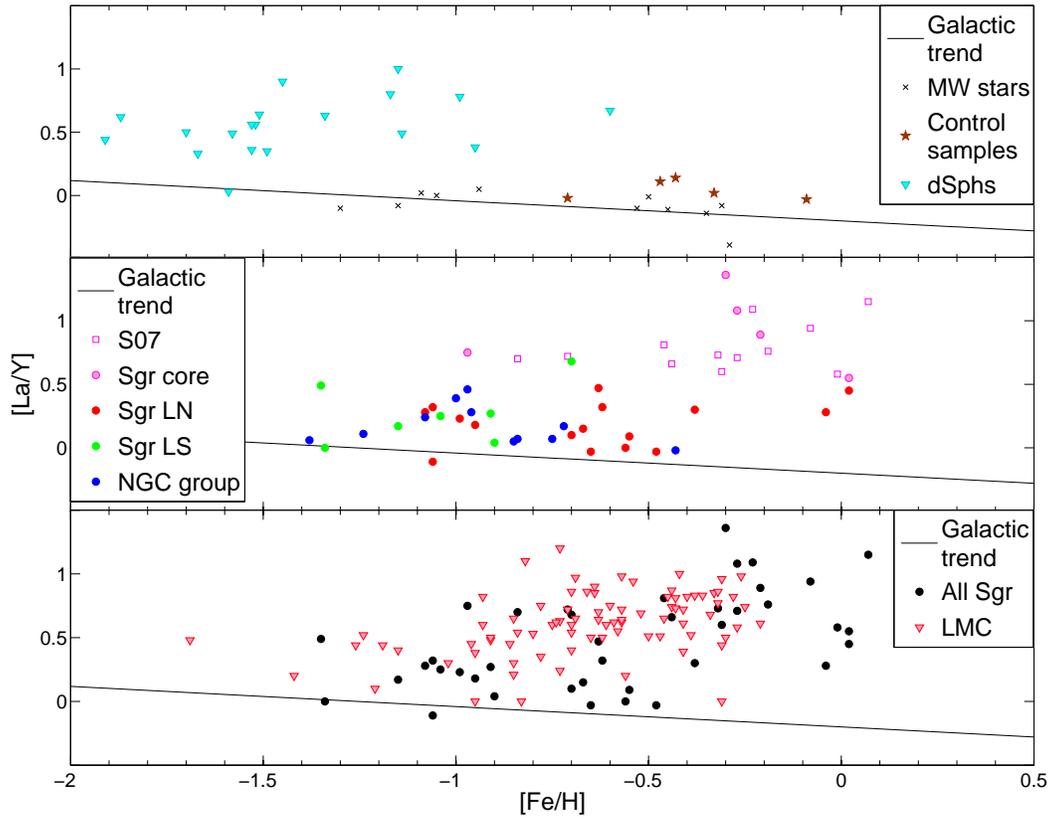} \caption{The same as Fig.\ 5, but for [La/Y]
versus [Fe/H].}
\end{figure}

\begin{figure}
\plotone{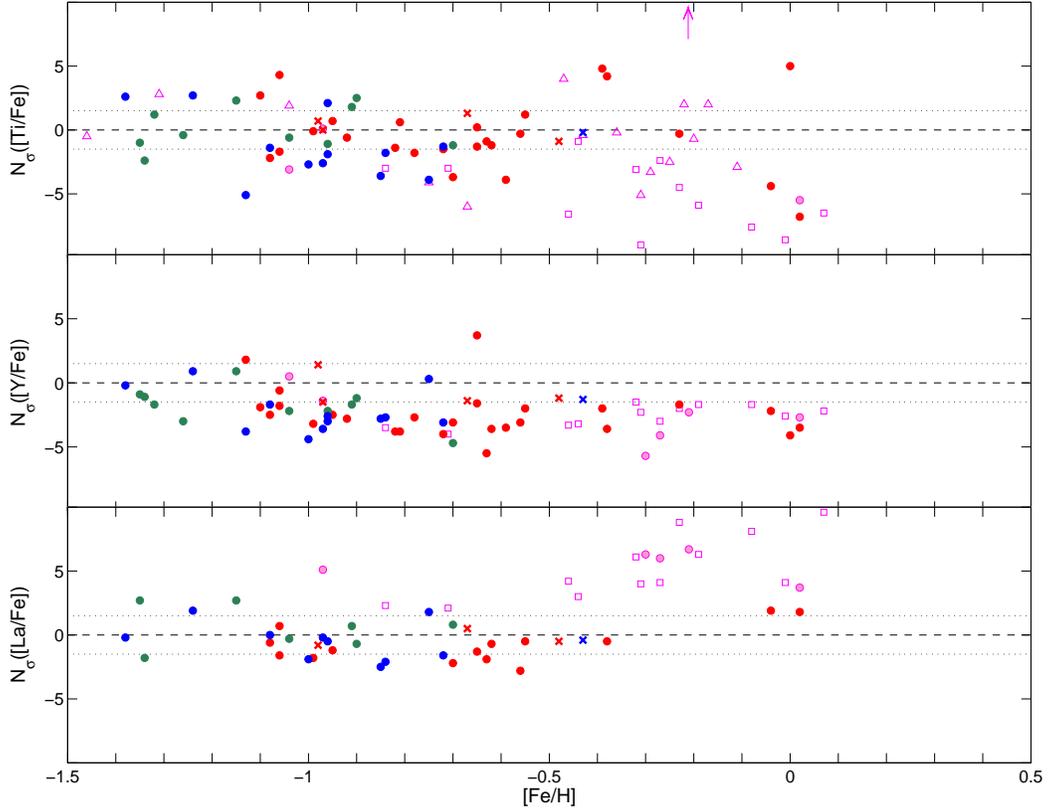} \caption{From top to bottom: N$_\sigma({\rm
[Ti/Fe]})$, N$_\sigma({\rm [Y/Fe]})$ and N$_\sigma({\rm [La/Fe]})$
versus [Fe/H], showing the quantitative deviation trend for each
subsample comparing with the Milky Way. The color coded as the
previous figures, the open symbols are Sgr core stars from Monaco
et al. (2005b) and Sbordone et al. (2007). The dashed lines
represent zero $\sigma$, and dotted lines are $\pm 1.5 \sigma$}
\end{figure}

\begin{figure}
\plotone{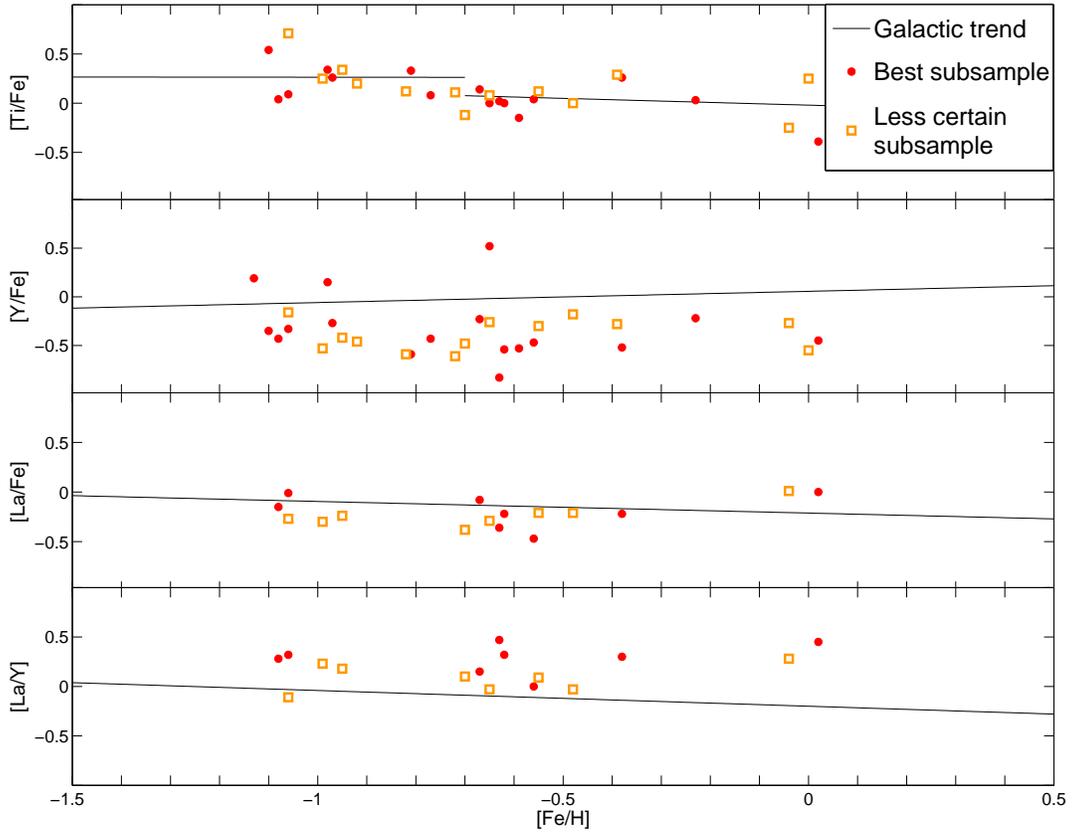} \caption{The distribution of [Ti/Fe], [Y/Fe],
[La/Fe] and [La/Y] as a function of [Fe/H] for the best and less
certain Sgr LN groups. }
\end{figure}

\begin{figure}
\includegraphics[scale=0.7] {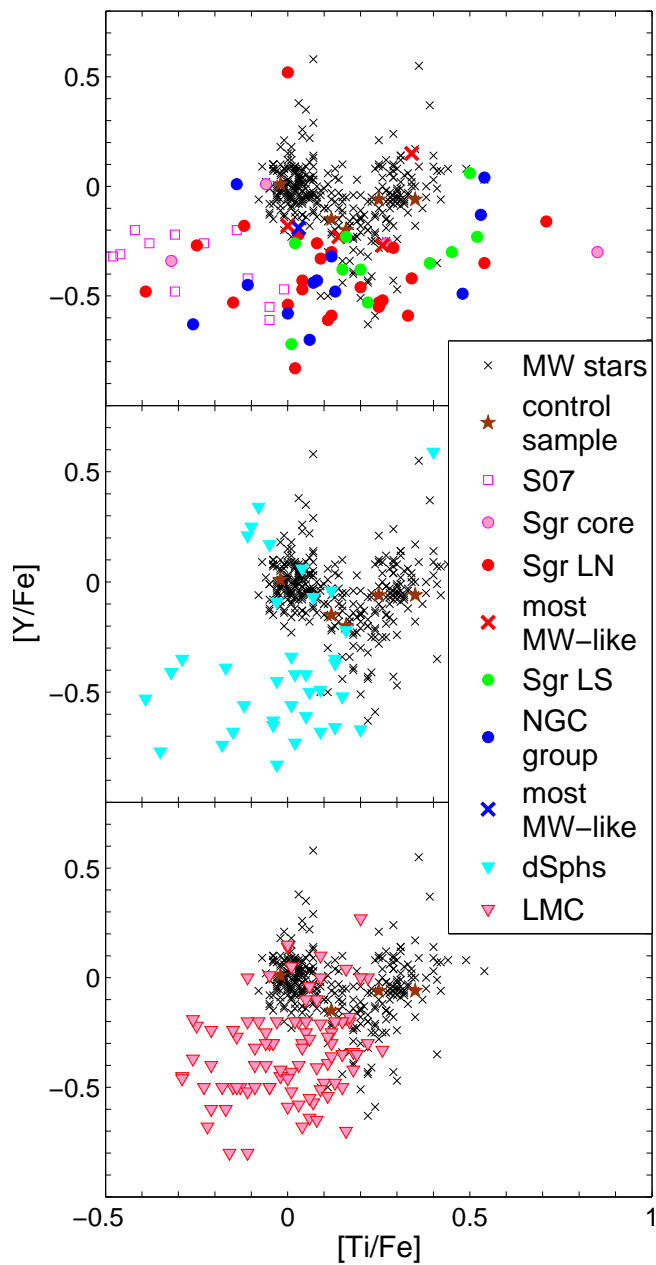}
\caption{ Distribution of [Y/Fe] versus [Ti/Fe] for our
stars ({\it top panel}), stars in other dSphs ({\it middle panel})
and stars in the LMC ({\it bottom panel}) --- with all stars color
coded as in the previous figures --- compared with Milky Way
stars. The latter lie within a well-defined, V-shaped distribution
near the top of each panel. }
\end{figure}

\begin{figure}
\plotone{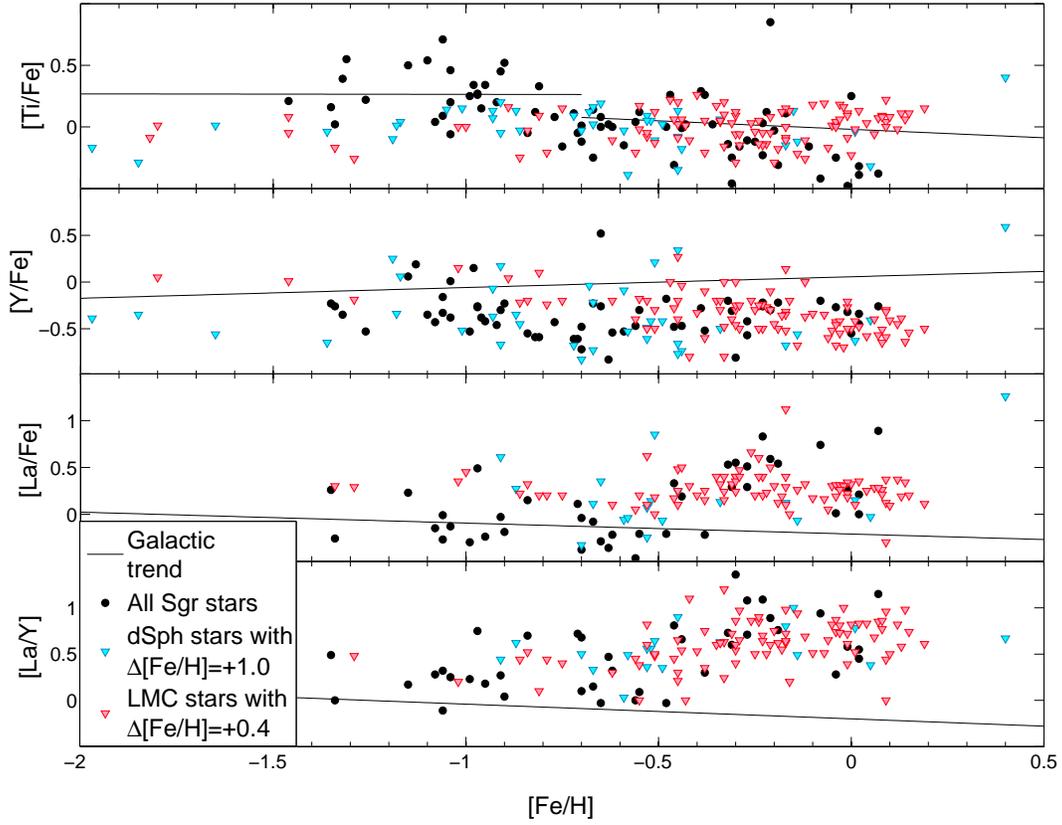} \caption{The distribution of [Ti/Fe], [Y/Fe],
[La/Fe] and [La/Y] as a function of [Fe/H] for Sgr, LMC and other
dSphs, but with a shift of +0.4 dex in [Fe/H] for LMC, and +1 dex
in [Fe/H] for other dSphs.}
\end{figure}

\end{document}